\documentclass{article}
\usepackage{amsfonts}
\usepackage{amsmath}
\usepackage{makeidx}
\usepackage{eurosym}
\usepackage{amssymb}
\usepackage{graphicx}
\usepackage{caption}
\usepackage{wrapfig}
\usepackage{subfig}
\usepackage{rotating}
\usepackage{natbib}

\newtheorem{theorem}{Theorem}
\newtheorem{acknowledgement}[theorem]{Acknowledgement}

\input{tcilatex}
\begin{document}

\title{Toward an efficient hybrid method for pricing barrier options on
assets with stochastic volatility }
\author{Alexander Lipton\thanks{%
Abu Dhabi Investment Authority, 211 Corniche, PO Box 3600, Abu Dhabi, United
Arab Emirates}, Artur Sepp\thanks{%
Sygnum Bank, Uetlibergstrasse 134a, 8045 Z\"{u}rich, Switzerland}}
\maketitle

\begin{abstract}
We combine the one-dimensional Monte Carlo simulation and the
semi-analytical one-dimensional heat potential method to design an efficient
technique for pricing barrier options on assets with correlated stochastic
volatility. Our approach to barrier options valuation utilizes two loops.
First we run the outer loop by generating volatility paths via the Monte
Carlo method. Second, we condition the price dynamics on a given volatility
path and apply the method of heat potentials to solve the conditional
problem in closed-form in the inner loop. We illustrate the accuracy and
efficacy of our semi-analytical approach by comparing it with the
two-dimensional Monte Carlo simulation and a hybrid method, which combines
the finite-difference technique for the inner loop and the Monte Carlo
simulation for the outer loop. We apply our method for computation of state
probabilities (Green function), survival probabilities, and values of call
options with barriers. Our approach provides better accuracy and is orders
of magnitude faster than the existing methods. s a by-product of our
analysis, we generalize Willard's (1997) conditioning formula for valuation
of path-independent options to path-dependent options and derive a novel
expression for the joint probability density for the value of drifted
Brownian motion and its running minimum.

\textbf{Keywords: }barrier options; stochastic volatility; Heston model;
heat potentials; semi-analytical solution; Volterra equation; Willards's
formula;

\textbf{2010 MSC:} 91G20; 91G60; 91G80; 47G10; 47G40; 35Q79;
\end{abstract}

\section{Introduction\label{Sec1}}

By expressing prices of European calls and puts in terms of the price of the
underlying asset and its volatility, \cite{Black1973} and \cite{Merton1973}
started the quantitative finance revolution. Their formula, which is known
as the Black-Scholes-Merton formula, is based on two assumptions: (a) the
risky price dynamics can be delta-hedged so that options can be valued using
the risk-neutral measure under which the underlying grows at the
risk-neutral rate; (b) price evolution is driven by a geometric Brownian
motion of the form:%
\begin{equation}
\frac{dS_{t}}{S_{t}}=rdt+\sigma dW_{t},\ \ \ S_{0}=s,  \label{Eq1}
\end{equation}%
where $W_{t}$ is a Brownian motion, $r$ and $\sigma $ are constant
risk-neutral interest rate and volatility of returns, respectively. Thus, at
time $T$, the price $S_{T}$ has the log-normal distribution:%
\begin{equation}
S_{T}=e^{\left( r-\frac{1}{2}\sigma ^{2}\right) T+\sigma W_{T}}S_{0},
\label{Eq2}
\end{equation}%
Under these assumptions, it is very easy to derive the price of, say, a call
option with maturity $T$ and strike $K$, with payoff of the form $\left(
S_{T}-K\right) _{+}$:%
\begin{equation}
\begin{array}{c}
C^{BS}(0,S_{0},T,K;r,\sigma )=S_{0}\mathfrak{N}\left( d_{+}\right) -e^{-rT}K%
\mathfrak{N}\left( d_{-}\right) , \\ 
d_{\pm }=\frac{-\ln \left( K/S_{0}\right) +rT\pm \sigma ^{2}T/2}{\sigma 
\sqrt{T}},%
\end{array}
\label{Eq3}
\end{equation}%
where $\mathfrak{N}(x)$ is the cdf for the standard $\left( 0,1\right) $
normal random variable. \cite{Lipton2002a} showed that in many situations it
is more convenient to write $C^{BS}$ as a Fourier integral:%
\begin{equation}
\begin{array}{c}
C^{BS}(0,S_{0},T,K;r,\sigma ) \\ 
=S_{0}\left( 1-\frac{1}{2\pi }\int\limits_{-\infty }^{\infty }\frac{%
e^{\left( i\chi +1/2\right) \left( \ln \left( K/S_{0}\right) -rT\right)
-\left( \chi ^{2}+1/4\right) \sigma ^{2}T/2}}{\left( \chi ^{2}+1/4\right) }%
d\chi \right) .%
\end{array}
\label{Eq3a}
\end{equation}

However, it became clear in a few years that the Black-Scholes formula,
taken literally, is not as helpful as initially thought since it could not
reproduce option market prices by using constant volatility for pricing
options with different strikes and maturities. The volatility skew (or
smile) effect (i.e., the need to use different volatilities to reproduce
market prices of European calls and puts with different strikes and
maturities) is observed in all markets. Therefore, it is of great interest
to practitioners and academics alike. Over the thirty years, many models
were developed to describe the smile effect.

Eventually, two approaches emerged - a reduced approach replacing the
constant volatility, central to the Black-Scholes approach, by the maturity-
and strike-dependent implied volatility, used to match the corresponding
market prices. Specifically, instead of a constant $\sigma $ for all $T,K$
in Eq. (\ref{Eq3}), a function $\sigma \left( T,K\right) $ is used. The
corresponding call price has the form%
\begin{equation}
C(0,S_{0},T,K)=C^{BS}(0,S_{0},T,K;r,\sigma \left( T,K\right) ).  \label{Eq4}
\end{equation}%
While useful in some situations, this approach is conceptually fairly
limited, since it does not explain why volatility is not constant. Hence, a
structural approach replacing the geometric Brownian motion with a
different, but still risk-neutral, driver for the underlying was quickly
developed. For instance, \cite{Derman1994}, \cite{Dupire1994}, and \cite%
{Rubinstein1994} simultaneously and independently developed the local
volatility model. \cite{Merton1976}, \cite{Andersen2000}, \cite{Lewis2001}
and many others developed jump-diffusion models. Stochastic volatility
models were developed by \cite{Hull1987}, \cite{Scott1987}, \cite%
{Wiggins1987}, \cite{Stein1991}, \cite{Heston1993}, \cite{Lewis2000}, \cite%
{Bergomi2015}, and others. Various combinations of the above were proposed,
for example by \cite{Dupire1996}, \cite{Jex1999}, \cite{Hagan2002}, \cite%
{Lipton2002a}, which culminated in Lipton's universal volatility model; see 
\cite{Lipton2002a}. The above-mentioned models are compared and contrasted
in \cite{Lipton2002a}.

\cite{Lipton2002b} explained that the actual worth of a structural model is
not in its ability to price vanilla options, which all structural models
worth their salt can do well, but in producing consistent prices for \emph{%
both} vanillas and first-generation exotics. Despite more than thirty years
of strenuous efforts, finding a proper theoretical framework and
implementing it in practice remains a significant challenge.

Pricing of exotic options in the presence of a smile is usually difficult
and seldom can be done analytically. Asymptotic methods developed by \cite%
{Hull1987}, \cite{Hagan2002}, \cite{Lipton1997}, \cite{Lipton2001}, among
others, proved to be very useful for solving the corresponding pricing
problems. Numerical methods, such as the Monte Carlo simulation (MCS) and
the finite difference method (FDM), are equally useful; see \cite%
{Glasserman2004}, \cite{Achdou2005}. Adding new methods to the classical
ones is definitely worth the effort. One can reduce many problems we wish to
solve to the initial-boundary value problems (IBVPs) for one-dimensional
parabolic partial differential equations (PDEs) with moving boundaries and
(or) time-dependent coefficients. Such problems appear naturally in various
areas of science and technology. Finding their semi-analytical solutions
requires using sophisticated tools, such as the method of heat potentials
(MHP) and a complementary method of generalized integral transforms (MGIT).
Such methods were actively developed by the Russian mathematical school in
the 20th century; see \cite{Kartashov2001} and references therein.

In the context of mathematical finance, A. Lipton and his coauthors actively
utilized the MHP; see \cite{Lipton2001}, \cite{Lipton2019} \cite{Lipton2020}
and references therein. In addition, A. Itkin and his coauthors used the
MGIT to price barrier and American options in the semi-closed form; see,
e.g., \cite{Carr2020}, \cite{Itkin2021a}, \cite{Carr2022}, \cite{Itkin2022}.
In principle, the MHP and MGIT can be generalized and used for any linear
differential operator with time-independent coefficients.

The MHP and MGIT boil down to solving linear Volterra equations of the
second kind and representing option prices as one-dimensional integrals. 
\cite{Itkin2021b} described the MHP and MGIT in the recent comprehensive
book and showed that they are much more efficient and provide better
accuracy and stability than the existing methods, such as the backward and
forward FDM or MCS.

This paper revisits the classical problem of pricing barrier options on
assets with stochastic volatility. We show that by using the concept of
conditional independence, we can reduce it to solving an initial-boundary
value problem with time-dependent coefficients and subsequent averaging over
the space of variance trajectories. Based on this observation, we develop an
efficient method that combines the MHP and MCS and provides a fast and
accurate solution to the problem at hand. Our method is very general and can
handle all known stochastic volatility models, as well as models with rough
volatility.\footnote{%
We intend to cover the latter topic in a separate publication.}

The paper is organized as follows. In Section \ref{Sec2}, we introduce
generic stochastic volatility models and describe the splitting method,
which allows one to study the dynamics of the log-price $X_{t}=\ln \left(
S_{t}/S_{0}\right) $, as a conditionally-independent one-dimensional
processes. We specialize these equations for the Heston and Stein-Stein
models. We also present the exact Lewis-Lipton and conditional Willard
formulas for vanilla options, such as European calls and puts on assets with
stochastic volatility and compare the corresponding prices. In Section \ref%
{Sec3}, we introduce barrier options on assets with stochastic volatility,
which are the main object of our study. We derive a conditional valuation
formula for such options, which generalizes the Willard formula for vanilla
options. In Section \ref{Sec4}, we describe a hybrid method for pricing
barrier options, which relies on the conditional independence decomposition.
The method consists of the outer Monte Carlo loop and the inner loop, which
requires solving the advection-diffusion problem for the drifted Brownian
motion with time-dependent coefficients on a semi-axis. We solve the latter
problem via two complementary methods: the FDM and the MHP. Results produced
by both methods are in perfect agreement. However, as expected, the second
method is orders of magnitude faster than the first one. In Section \ref%
{Sec4}, we use the MHP to solve an old problem in probability theory and
show how to find the joint probability density for the value of drifted
Brownian motion and its running minimum via the MHP. We draw our conclusions
in Section \ref{Sec6}.

\section{Stochastic volatility models\label{Sec2}}

\subsection{Conditionally independent dynamics\label{Sec21}}

We consider the joint evolution of the asset price, $S_{t}$, and its
stochastic variance, $V_{t}$, as follows: 
\begin{equation}
\begin{array}{c}
\frac{dS_{t}}{S_{t}}=rdt+\sqrt{V_{t}}\left( \rho dB_{t}+\sqrt{1-\rho ^{2}}%
dW_{t}\right) ,\ \ \ S_{0}=s, \\ 
dV_{t}=\Phi \left( V_{t}\right) dt+\Psi \left( V_{t}\right) dB_{t},\ \ \
V_{0}=v,%
\end{array}
\label{Eq5}
\end{equation}%
where $B_{t}$ and $W_{t}$ are independent Brownian motions. Given that Eqs (%
\ref{Eq5}) are scale invariant with respect to $S_{t}$, we can write them in
terms of $X_{t}=\ln \left( S_{t}/S_{0}\right) $ and $V_{t}$:%
\begin{equation}
\begin{array}{c}
dX_{t}=\left( r-\frac{1}{2}V_{t}\right) dt+\sqrt{V_{t}}\left( \rho dB_{t}+%
\sqrt{1-\rho ^{2}}dW_{t}\right) ,\ \ \ \ X_{0}=0, \\ 
dV_{t}=\Phi \left( V_{t}\right) dt+\Psi \left( V_{t}\right) dB_{t},\ \ \
V_{0}=v,%
\end{array}
\label{Eq7}
\end{equation}%
Alternatively, we can study the joint evolution of the price $S_{t}$ and its
volatility $\sigma _{t}$:%
\begin{equation}
\begin{array}{c}
\frac{dS_{t}}{S_{t}}=rdt+\sigma _{t}\left( \rho dB_{t}+\sqrt{1-\rho ^{2}}%
dW_{t}\right) ,\ \ \ S_{0}=s, \\ 
d\sigma _{t}=\phi \left( \sigma _{t}\right) dt+\psi \left( \sigma
_{t}\right) dB_{t},\ \ \ \sigma _{0}=\sigma .%
\end{array}
\label{Eq6}
\end{equation}%
Given that Eqs (\ref{Eq6}) are scale invariant with respect to $S_{t}$, we
can write them in terms of $X_{t}=\ln \left( S_{t}/S_{0}\right) $ and $%
\sigma _{t}$:%
\begin{equation}
\begin{array}{c}
dX_{t}=\left( r-\frac{1}{2}\sigma _{t}^{2}\right) dt+\sigma _{t}\left( \rho
dB_{t}+\sqrt{1-\rho ^{2}}dW_{t}\right) ,\ \ \ X_{0}=0, \\ 
d\sigma _{t}=\phi \left( \sigma _{t}\right) dt+\psi \left( \sigma
_{t}\right) dB_{t},\ \ \ \sigma _{0}=\sigma ,%
\end{array}
\label{Eq8}
\end{equation}%
which is often more convenient. From now on, we shall concentrate of
studying the dynamics of $X_{t}$.

In the general case, we can write $dB_{t}$ in the form%
\begin{equation}
dB_{t}=\frac{dV_{t}-\Phi \left( V_{t}\right) dt}{\Psi \left( V_{t}\right) },
\label{Eq9}
\end{equation}%
and obtain the following conditionally-independent dynamics for the
log-price $X_{t}$:%
\begin{equation}
dX_{t}=\left( r-\frac{V_{t}}{2}-\frac{\rho \sqrt{V_{t}}\Phi \left(
V_{t}\right) }{\Psi \left( V_{t}\right) }\right) dt+\frac{\rho \sqrt{V_{t}}%
dV_{t}}{\Psi \left( V_{t}\right) }+\sqrt{1-\rho ^{2}}\sqrt{V_{t}}dW_{t},\ \
\ X_{0}=0.  \label{Eq10}
\end{equation}%
Similarly, we can write $dB_{t}$ in the form%
\begin{equation}
dB_{t}=\frac{d\sigma _{t}-\phi \left( \sigma _{t}\right) }{\psi \left(
\sigma _{t}\right) },  \label{Eq11}
\end{equation}%
and get the following dynamics for $X_{t}$:%
\begin{equation}
dX_{t}=\left( r-\frac{1}{2}\sigma _{t}^{2}-\frac{\rho \sigma \phi \left(
\sigma _{t}\right) }{\psi \left( \sigma _{t}\right) }\right) dt+\frac{\rho
\sigma _{t}d\sigma _{t}}{\psi \left( \sigma _{t}\right) }+\sqrt{1-\rho ^{2}}%
\sigma _{t}dW_{t},\ \ \ X_{0}=0.  \label{Eq12}
\end{equation}%
Assuming that the variance or volatility paths are given, Eqs (\ref{Eq10}), (%
\ref{Eq12}) describe drifted arithmetic Brownian motion with time-dependent
drift and volatility.

For the well-known Heston model, \cite{Heston1993}, we have%
\begin{equation}
\begin{array}{c}
\Phi \left( V_{t}\right) =\kappa \left( \theta -V_{t}\right) ,\ \ \ \Psi
\left( V_{t}\right) =\varepsilon \sqrt{V_{t}}, \\ 
dV_{t}=\kappa \left( \theta -V_{t}\right) dt+\varepsilon \sqrt{V_{t}}dB_{t},%
\end{array}
\label{Eq13}
\end{equation}%
so that Eq. (\ref{Eq10}) has the form%
\begin{equation}
dX_{t}=\left( r-\frac{\rho \kappa \theta }{\varepsilon }-\left( \frac{1}{2}-%
\frac{\rho \kappa }{\varepsilon }\right) V_{t}\right) dt+\frac{\rho }{%
\varepsilon }dV_{t}+\sqrt{1-\rho ^{2}}\sqrt{V_{t}}dW_{t}.  \label{Eq15}
\end{equation}%
Thus, $X_{t}$ is the so-called drifted Brownian motion driven by the
stochastic differential equation of the form%
\begin{equation}
\begin{array}{c}
dX_{t}=\mu \left( t\right) dt+\nu \left( t\right) dW_{t}, \\ 
\mu \left( t\right) =\left( r-\frac{\rho \kappa \theta }{\varepsilon }%
-\left( \frac{1}{2}-\frac{\rho \kappa }{\varepsilon }\right) V_{t}\right) +%
\frac{\rho }{\varepsilon }\frac{dV_{t}}{dt}, \\ 
\nu \left( t\right) =\sqrt{1-\rho ^{2}}\sqrt{V_{t}}.%
\end{array}
\label{Eq15a}
\end{equation}

For the Stein-Stein model, \cite{Stein1991}, we have%
\begin{equation}
\begin{array}{c}
\phi \left( \sigma _{t}\right) =\hat{\kappa}\left( \hat{\theta}-\sigma
_{t}\right) ,\ \ \ \psi \left( \sigma _{t}\right) =\hat{\varepsilon}, \\ 
d\sigma _{t}=\hat{\kappa}\left( \hat{\theta}-\sigma _{t}\right) dt+\hat{%
\varepsilon}dB_{t},%
\end{array}
\label{Eq14}
\end{equation}%
so that Eq. (\ref{Eq12}) becomes 
\begin{equation}
dX_{t}=\left( r-\frac{\rho \hat{\kappa}\hat{\theta}}{\hat{\varepsilon}}%
\sigma _{t}-\left( \frac{1}{2}-\frac{\rho \hat{\kappa}}{\hat{\varepsilon}}%
\right) \sigma _{t}^{2}\right) dt+\frac{\rho }{\hat{\varepsilon}}\sigma
_{t}d\sigma _{t}+\sqrt{1-\rho ^{2}}\sigma _{t}dW_{t}.  \label{Eq16}
\end{equation}

Traditionally, the Heston model is viewed as better describing the market
than the Stein-Stein model since the latter does allow for zero variance. In
our opinion, this is not a particularly important issue, outweighed by many
advantages of the Stein-Stein model, such as a very straightforward way of
simulating the evolution of the volatility path. The Heston model gained
popularity due to the simple fact that it is exactly solvable for vanilla
options. The explicit solution can be calculated via the original \cite%
{Heston1993} formula. However, the Lewis-Lipton formula is much more
efficient; see \cite{Lewis2000}, \cite{Lipton2001}, \cite{Lipton2002a}, \cite%
{Lipton2008}, and \cite{Schmeltze2010}.

In this paper, we consider the Heston model with constant coefficients to
follow a long-established tradition, even though it is not necessarily our
preferred model. We emphasize that our method is general and can handle any
reasonable stochastic volatility model.

\subsection{Analytical valuation formula for vanilla options\label{Sec22}}

The popularity of the Heston model stems from the fact that one can write
its solution in the closed-form; see \cite{Heston1993}. However, experience
has shown that using the original formula is difficult due to several
technical drawbacks. Therefore, for benchmarking purposes, here we present
the Lewis-Lipton formula, which is easy to implement and use in practice:%
\begin{equation}
\begin{array}{c}
C^{H}\left( 0,S_{0},K,T;r,\rho ,\kappa ,\theta ,\varepsilon ,V_{0}\right) \\ 
=S_{0}\left( 1-\frac{1}{2\pi }\int\limits_{-\infty }^{\infty }\frac{%
e^{\left( i\chi +1/2\right) \left( \ln \left( K/S_{0}\right) -rT\right)
+\alpha \left( T,\chi \right) -\left( \chi ^{2}+1/4\right) \beta \left(
T,\chi \right) V_{0}}}{\left( \chi ^{2}+1/4\right) }d\chi \right) , \\ 
\alpha \left( T,\chi \right) =-\frac{\kappa \theta }{\varepsilon ^{2}}\left[
\psi _{+}T+2\ln \left( \frac{\psi _{-}+\psi _{+}\exp \left( -\zeta T\right) 
}{2\zeta }\right) \right] , \\ 
\beta \left( T,\chi \right) =\frac{1-\exp \left( -\zeta T\right) }{\psi
_{-}+\psi _{+}\exp \left( -\zeta T\right) }, \\ 
\psi _{\pm }=\mp \left( i\rho \varepsilon \chi +\hat{\kappa}\right) +\zeta ,
\\ 
\zeta =\sqrt{\varepsilon ^{2}\left( 1-\rho ^{2}\right) \chi
^{2}+2i\varepsilon \rho \hat{\kappa}\chi +\hat{\kappa}^{2}+\frac{\varepsilon
^{2}}{4}},%
\end{array}
\label{Eq16a}
\end{equation}%
where $\hat{\kappa}=\kappa -\rho \varepsilon /2$. Further details are given
in \cite{Lewis2000}, \cite{Lipton2001}, \cite{Lipton2002a}, and \cite%
{Schmeltze2010}.\footnote{%
There is a typo in \cite{Lipton2002a} - a minus sign in front of $\beta $.
This typo is corrected in \cite{Lipton2018}, Chapter 10.} It is clear that
Eq. (\ref{Eq16a}) is a generalization of Eq. (\ref{Eq3a}).

\subsection{Conditional valuation formula for vanilla options\label{Sec23}}

Unfortunately, with very few exceptions, finding a closed-form solution for
barrier or other exotic options on assets with stochastic volatility is not
possible, even if such a solution exists for vanilla options. Hence, more
general volatility models for barrier options are as good (or bad) as the
more traditional Heston and Stein-Stein models, which enjoy closed-form
solutions for vanilla options.

We express the log-return process as a linear combination of the two
processes: 
\begin{equation}
X_{t}=Y_{t}+\left( \left( r-\frac{\rho \kappa \theta }{\varepsilon }\right)
t-\left( \frac{1}{2}-\frac{\rho \kappa }{\varepsilon }\right) I_{t}\right) +%
\frac{\rho }{\varepsilon }\left( V_{t}-V_{0}\right) \equiv Y_{t}+M_{t},
\label{Eq20}
\end{equation}%
where 
\begin{equation}
\begin{array}{c}
Y_{t}=\sqrt{1-\rho ^{2}}\int_{0}^{t}\sqrt{V_{t}}dW_{t},\ Y_{0}=0, \\ 
I_{t}=\int_{0}^{t}V_{t^{\prime }}dt^{\prime },\ I_{0}=0, \\ 
M_{t}=\left( \left( r-\frac{\rho \kappa \theta }{\varepsilon }\right)
t-\left( \frac{1}{2}-\frac{\rho \kappa }{\varepsilon }\right) I_{t}\right) +%
\frac{\rho }{\varepsilon }\left( V_{t}-V_{0}\right) ,\ \ \ M_{0}=0.%
\end{array}
\label{Eq21}
\end{equation}%
Accordingly, conditionally on the filtration generated by the variance
process $V_{t}$, we can represent the solution to the price process given by
Eq. (\ref{Eq5}) as follows: 
\begin{equation}
S_{t}=e^{M_{t}+Y_{t}}S_{0},  \label{Eq22}
\end{equation}%
where $M_{t}$ is interpreted as a time-deterministic cumulative drift, and $%
Y_{t}$ is a martingale with a deterministic time-dependent quadratic
variance.

It is clear that pricing for path-independent options, such as European
calls and puts, simplifies to the Black-Scholes-Merton formula, provided
that either a variance path $\left\{ \left. V_{t}\right\vert 0\leq t\leq
T\right\} $ (or a volatility path $\left\{ \left. \sigma _{t}\right\vert
0\leq t\leq T\right\} $) is given. To be concrete, consider a call option
with maturity $T$ and strike $K$. Eq. (\ref{Eq5}) yields 
\begin{equation}
\frac{dS_{t}}{S_{t}}=\left( \left( r-\frac{1}{2}V_{t}\right) dt+\rho \sqrt{%
V_{t}}dB_{t}\right) +\sqrt{1-\rho ^{2}}\sqrt{V_{t}}dW_{t},\ \ \ ,\ \ S_{0}=s.
\label{Eq23}
\end{equation}%
Thus,%
\begin{equation}
S_{T}=e^{rT-\frac{1}{2}\left( 1-\rho ^{2}\right) I_{T}+\sqrt{1-\rho ^{2}}%
\sqrt{I_{T}}\eta }\left( e^{-\frac{1}{2}\rho ^{2}I_{T}+\rho
J_{T}}S_{0}\right) ,  \label{Eq24}
\end{equation}%
where the non-dimensional random variables $I_{T},J_{T}$, are given by%
\begin{equation}
I_{T}=\int_{0}^{T}V_{t}dt,\ \ \ J_{T}=\int_{0}^{T}\sqrt{V_{t}}dB_{t},
\label{Eq25}
\end{equation}%
and $\eta $ is the standard $\left( 0,1\right) $ normal variable.
Accordingly for a particular trajectory $\left\{ \left. V_{t}\right\vert
0\leq t\leq T\right\} $, we obtain the following expression%
\begin{equation}
C=C^{BS}\left( 0,e^{-\frac{1}{2}\rho ^{2}I_{T}+\rho J_{T}}S_{0},T,K;r,\sqrt{%
1-\rho ^{2}}\sqrt{\frac{I_{T}}{T}}\right) ,  \label{Eq26}
\end{equation}%
where the values $\left( I_{T},J_{T}\right) $ are assumed to be known. The
unconditional price is obtained by averaging over all possible $\left(
I_{T},J_{T}\right) $:%
\begin{equation}
\begin{array}{c}
C^{H}\left( 0,S_{0},K,T;r,\rho ,\kappa ,\theta ,\varepsilon ,v_{0}\right) \\ 
=\int_{0}^{\infty }\int_{0}^{\infty }C^{BS}\left( 0,e^{-\frac{1}{2}\rho
^{2}I_{T}+\rho J_{T}}S_{0},T,K;r,\sqrt{1-\rho ^{2}}\sqrt{\frac{I}{T}}\right)
\Xi \left( I_{T},J_{T}\right) dJ_{T}dI_{T},%
\end{array}
\label{Eq27}
\end{equation}%
where $\Xi \left( I_{T},J_{T}\right) $ is the joint probability density
function (pdf) for the pair $\left( I_{T},J_{T}\right) $; see \cite%
{Willard1997} and \cite{Romano1997}. Thus, Eq. (\ref{Eq27}) splits the
calculation of the call option price into two stages. First, the conditional
price is found analytically via the standard Black-Scholes formula.\ Second,
this conditional price is averaged according to the particular choice of the
process for the variance $V_{t}$. Of course, the first stage is trivial. The
usefulness of this formula depends on how easy (or difficult) it is to find
the pdf for $\left( I,J\right) $ and complete the second stage.

Two approaches have been used in practice - the standard Monte-Carlo method
for calculating $\left\{ \left. V_{t}\right\vert 0\leq t\leq T\right\} $, $I$%
, $J$, and a more advanced (but much harder) method based on the
augmentation principle described in Section (13.2) \cite{Lipton2001}.
Specifically, the augmented dynamic equation for $V_{t}$ yields the
following system of degenerate PDEs for the triple $\left( V,I,J\right) $:%
\begin{equation}
\begin{array}{c}
dV_{t}=\Phi \left( V_{t}\right) dt+\Psi \left( V_{t}\right) dB_{t},\ \ \
V_{0}=v, \\ 
dI_{t}=V_{t}dt,\ \ \ I_{0}=0, \\ 
dJ_{t}=\sqrt{V_{t}}dB_{t},\ \ \ J_{0}=0.%
\end{array}
\label{Eq28}
\end{equation}%
The corresponding Green's function $G\left( t,V,I,J\right) $ is governed by
the degenerate Fokker-Planck equation of the form%
\begin{equation}
\begin{array}{c}
G_{t}+\left( \Phi \left( V\right) G\right) _{V}+VG_{I}-\frac{1}{2}\left(
\Psi ^{2}\left( V\right) G\right) _{VV}-\left( \sqrt{V}\Psi \left( V\right)
G\right) _{VJ}-\frac{1}{2}\left( VG\right) _{JJ}=0, \\ 
G\left( 0,V,I,J\right) =\delta \left( V-v\right) \delta \left( I\right)
\delta \left( J\right) .%
\end{array}
\label{Eq29}
\end{equation}%
In general, solving this equation is complicated; however, for the Heston
model, \cite{Lipton2001} found a closed-form solution.

\begin{figure}[tbp]
\begin{center}
{%
\includegraphics[width=0.8\textwidth]
{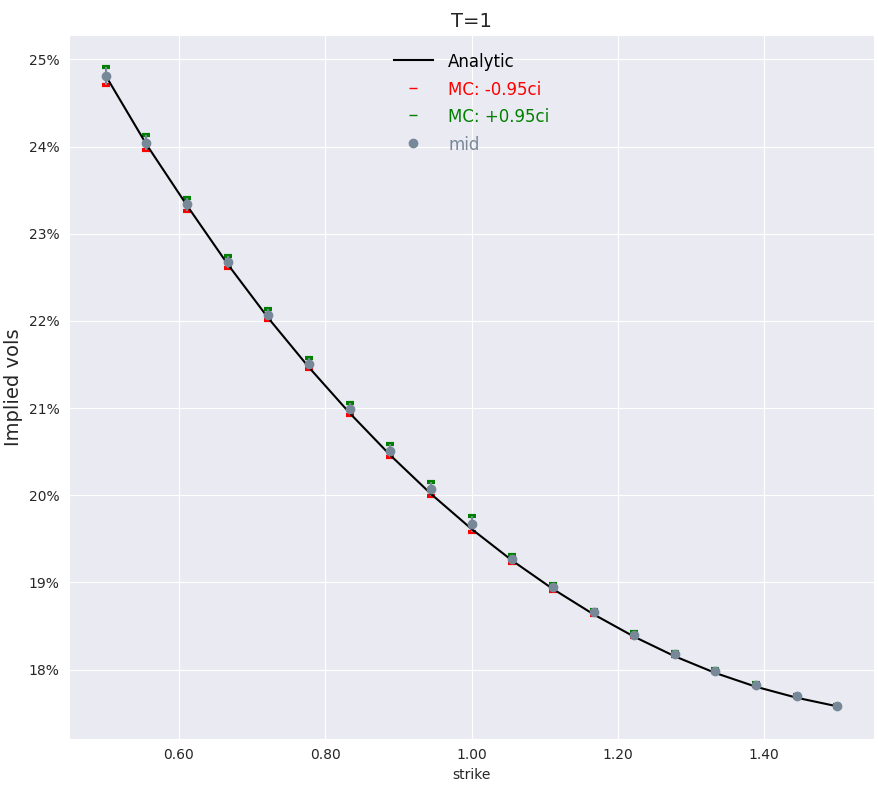}}
\end{center}
\par
\vspace{-10pt}
\caption{Implied volatilities of European call options. We obtain the
corresponding prices by using Eqs. (\protect\ref{Eq16a}) and (\protect\ref%
{Eq27}). Here, and throughout the paper we use the following parameters: $%
S_{0}=1$, $V_{0}=0.25$, $T=1.0$, $r=0.03$, $\protect\kappa =1.0$, $\protect%
\theta =0.2$, $\protect\rho =-0.3$, $\protect\varepsilon =0.4$.}
\label{Fig1}
\end{figure}

In Figure \ref{Fig1} we compare prices of European call options given by
analytical formula (\ref{Eq16a}), and Monte Carlo formula (\ref{Eq27}). As
expected, both formulas agree, although the latter formula is much slower
than the former. 

\section{Barrier options\label{Sec3}}

\subsection{Formulation\label{Sec31}}

Our task is to price a barrier option written on an asset with stochastic
volatility. For brevity, we consider barrier options with the lower barrier $%
B<S_{0}$ only. Considering other possibilities, such as pricing options with
the upper barrier or popular double-no-touch options, is left to the reader.
The corresponding IBVP can be written in the form%
\begin{equation}
\begin{array}{c}
P_{t}+\left( r-\frac{1}{2}V\right) P_{X}+\kappa \left( \theta -V\right)
P_{V}+\frac{1}{2}V\left( P_{XX}+2\rho P_{XV}+P_{VV}\right) -rP=0, \\ 
P\left( T,X,V\right) =\Pi \left( X\right) , \\ 
P\left( t,\xi ,V\right) =0,%
\end{array}
\label{Eq17}
\end{equation}%
where $\xi =\ln \left( B/S_{0}\right) <0$, $\Pi (X)$ is the terminal payoff
function. Typical examples include the no-touch, call and put payoffs
defined by: 
\begin{equation}
\Pi \left( X_{T}\right) =1,\ \Pi \left( X_{T}\right) =S_{0}\left(
e^{X_{T}}-e^{k}\right) _{+},\ \Pi \left( X_{T}\right) =S_{0}\left(
e^{k}-e^{X_{T}}\right) _{+},  \label{Eq33}
\end{equation}%
where $k=\ln \left( K/S_{0}\right) $. Alternatively, we can write the
adjoint problem for Green's function $G$:%
\begin{equation}
\begin{array}{c}
G_{t}+\left( \left( r-\frac{1}{2}V\right) G\right) _{X}+\left( \kappa \left(
\theta -V\right) G\right) _{V} \\ 
-\frac{1}{2}\left( \left( VG\right) _{XX}+2\rho \left( VG\right)
_{XV}+\left( VG\right) _{VV}\right) +rG=0, \\ 
G\left( 0,X,V\right) =\delta \left( X\right) \delta \left( V-V_{0}\right) ,
\\ 
G\left( t,\xi ,V\right) =0,%
\end{array}
\label{Eq18}
\end{equation}%
Once the corresponding Green's function is calculated, we can find $P$ via
simple integration:%
\begin{equation}
P\left( 0,0,V_{0}\right) =\int_{\xi }^{\infty }\int_{0}^{\infty }G\left(
0,0,V_{0},T,X_{T},V_{T}\right) \Pi \left( X_{T}\right) dV_{T}dX_{T}.
\label{Eq19}
\end{equation}%
While such options can be priced via either FDM or MCS, both are notoriously
slow. Therefore, we want to design a much faster method, enjoying equal or
higher accuracy than the classical alternatives.

As far as analytical solutions are concerned, only one is known. It was
discovered by \cite{Lipton1997}, see also \cite{Lipton2001}, \cite%
{Lipton2002b}, and \cite{Andreasen2001}. \cite{Lipton1997} observed that in
the special case $r=0$, $\rho =0$, IBVPs (\ref{Eq17}), (\ref{Eq18}) are
symmetric with respect to the transformation $X\rightarrow -X$. Hence, the
classical method of images is applicable, and solutions to these problems
can be presented as a linear combination of solutions without barriers. Of
course, one can use this approach for options in the presence of an upper
barrier, as well as for double-barrier options.

Recently, there were several unsuccessful attempts to solve the pricing
problem with $r^{2}+\rho ^{2}>0$. For example, \cite{Aquino2019} presented a
solution,\ relying on an explicit expression for the joint distribution of
the value of a Brownian motion with time-dependent drift and its maximum and
minimum; it was quickly shown by one of the present authors that their
calculation is erroneous; see \cite{Aquino2021}.\footnote{%
Finding the joint distribution for a Brownian motion with time-dependent
drift and its maximum and minimum is challenging. We present its solution in
Section \ref{Sec5}.} \cite{He2021} presented a \textquotedblleft
solution\textquotedblright ,\ which relies on the unsubstantiated
replacement of the time-dependent drift by a constant. Their approach is so
arbitrary and frivolous that its detailed repudiation is not warranted.

\subsection{Conditional valuation formula for barrier options\label{Sec32}}

It is hard to extend interesting formula (\ref{Eq27}) for barrier options.
However, it is not impossible! Following Section13.3 in \cite{Lipton2001},
we express the value of a path-dependent option as an integral in the
functional space of price trajectories: 
\begin{equation}
P(S_{0})=e^{-rT}\int_{\Omega }\mathcal{F}(\omega )d\mathcal{D}(\omega )
\label{Eq30}
\end{equation}%
where $\mathcal{F}(\omega )$ is a functional mapping of the space of
trajectories into payoffs, and $\mathcal{D}(\omega )$ is the risk-neutral
probability measure.

Further, by applying the augmentation principle, we introduce the functional 
$\Lambda _{t}$ to represent the path-dependent variable linked to the
evolution of the spot price $S_{t}$. We then consider evaluation of a
derivatives security with the terminal pay-off function $f(S,\Lambda )$.
Finally, we extend the joint dynamics in Eq. (\ref{Eq5}) with the dynamics
of augmented variable $\Lambda _{t}=\min_{0\leq t^{\prime }\leq t}S_{t}$:%
\begin{equation}
\begin{array}{c}
S_{t}=rS_{t}dt+\sqrt{V_{t}}S_{t}\left( \rho dB_{t}+\sqrt{1-\rho ^{2}}%
dW_{t}\right) ,\ \ \ S_{0}=s, \\ 
dV_{t}=\Phi (V_{t})dt+\Psi (V_{t})dB_{t},\ \ \ V_{0}=v, \\ 
d\Lambda _{t}=\theta \left( \Lambda _{t}-S_{t}\right) \left( dS_{t}\right)
_{-},\ \ \ \Lambda _{0}=S_{0}.%
\end{array}
\label{Eq31}
\end{equation}%
The payoff function $f(S,\Lambda )$ is given by: 
\begin{equation}
f(S_{T},\Lambda _{T})=\mathbf{1}_{\{\Lambda _{T}>B\}}\Pi \left( S_{T}\right)
\label{Eq32}
\end{equation}

The joint dynamic, given by Eqs (\ref{Eq31}), is Markovian. Accordingly, we
can reduce the general formula (\ref{Eq30}) to the form: 
\begin{equation}
P(0,S_{0})=e^{-rT}\int_{V=0}^{\infty }\int_{S=0}^{\infty }\int_{\Lambda
=0}^{\infty }f(S,\Lambda )G(T,V,S,\Lambda ;0,V_{0},S_{0},\Lambda
_{0})d\Lambda dSdV.  \label{Eq34}
\end{equation}%
Here $G(T,V,S,\Lambda ;0,V_{0},S_{0},\Lambda _{0})$ is the risk-neutral
probability density function for the joint evolution of state variables in
SDE (\ref{Eq31}). We emphasize that while the payoff function does not
depend on the variance $V$, the valuation problem has three spatial
variables, including variance, in addition to the time variable.

Below, we introduce a novel method to solve the valuation problem (\ref{Eq34}%
) semi-analytically. Using Eq. (\ref{Eq22}), the density of the price $S_{t}$
is log-normal with time-dependent drift and variance conditional on the
filtration generated by stochastic variance $V_{t}$, $\mathbb{F}^{V}$.
Therefore, we represent the valuation formula (\ref{Eq34}) as follows: 
\begin{equation}
P(0,S_{0})=e^{-rT}\mathbb{E}_{V}\left[ \int_{S=0}^{\infty }\int_{\Lambda
=0}^{\infty }f(S,\Lambda )\Gamma \left( \left. T,S,\Lambda ;0,S_{0},\Lambda
_{0}\right\vert V(\omega )\right) d\Lambda dS\right] ,  \label{Eq35}
\end{equation}%
where $\Gamma $ is the Gaussian density of $S_{t}$ and $\Lambda _{t}$
conditional of the variance path $V(\omega )$, and the expectation is
computed over all paths of variance process $V_{t}$. The term in the square
brackets is the value of a path-dependent option on $\Lambda $ and $S$,
computed in closed-form or numerically.

We apply the MHP to compute the inner integral for barrier options in the
closed-form. Thus, we have generalized Eq (\ref{Eq27}), valid for
path-independent options, to path-dependent options, and apply the new
result to value barrier options in the semi-closed form.

\section{Mixed MHP-MCS approach to solving IBVPs\label{Sec4}}

We consider the IBVP (\ref{Eq17}). In the spirit of Section \ref{Sec32}, we
split our algorithm in two steps: the outer MCS loop, which generates a
bunch of trajectories for the stochastic variance $v_{t}$, and the inner
loop, which solves the one-dimensional IBVP for $X_{t}$. We discuss two
approaches for solving the latter problem: the finite difference approach
developed by \cite{Loeper2009}, and the MHP approach inspired by \cite%
{Lipton2019}, \cite{Lipton2020}. We start with the inner loop. The outer
loop is relatively straightforward. It requires to average the inner price
by using the equation for $V_{t}$. This will provide solution of the form
given by Eq.\ (2) in \cite{Lipton2002b}. The simplest way of doing this
averaging is via MCS, although in some exceptional cases other approaches
can be envisaged.

\subsection{One-dimensional IBVP\label{Sec41}}

Eq. (\ref{Eq15}), conditional on the variance path, can be written as an SDE
of the form%
\begin{equation}
dX_{t}=\mu \left( t\right) dt+\nu \left( t\right) dW_{t},\ \ \ X_{t}\geq \xi
,  \label{Eq36}
\end{equation}%
where $X_{t}$ the drifted Brownian motion with $\mu ,\nu $ given by Eqs (\ref%
{Eq15a}).

While studying $X_{t}$ on the entire axis $\left( -\infty ,\infty \right) $
is almost trivial, dealing with the same process in a semi-bounded domain $%
\left( \xi ,\infty \right) $ might be quite hard. This paper shows how to do
it by using the FDM and the MHP.

We can always rescale time $t$ by introducing $\upsilon $, such that $%
d\upsilon =\nu \left( t\right) dt$. As a result, the governing SDE becomes%
\begin{equation}
dX_{\upsilon }=\lambda \left( \upsilon \right) d\upsilon +dW_{\upsilon },
\label{Eq37}
\end{equation}%
where $\lambda \left( \upsilon \right) =\mu \left( t\right) /\nu \left(
t\right) $, $\upsilon \left( t\right) =\int_{0}^{t}v\left( t^{\prime
}\right) dt^{\prime }$.

Using the decomposition for $X_{t}$ given by Eq. (\ref{Eq20}) we present the
valuation problems follows: 
\begin{equation}
dX_{t}=dY_{t}+dM_{t},\ \ \ Y_{0}=0,\ \ \ M_{0}=0,\ \ \ X_{t}\geq \xi ,
\label{Eq38}
\end{equation}%
where $Y_{t}$ is stochastic part and $M_{t}$ is deterministic part, and $\xi
<0$. Then we can express the problem (\ref{Eq20}) in terms of stochastic
variable $Y_{t}$ as follows: 
\begin{equation}
dY_{t}=\sqrt{1-\rho ^{2}}\sqrt{V_{t}}dW_{t},\ \ \ Y_{0}=0,\ \ \ Y_{t}\geq
\xi -M_{t}.  \label{Eq39}
\end{equation}%
We use a new time variable%
\begin{equation}
\upsilon \left( t\right) =\left( 1-\rho ^{2}\right) I_{t},  \label{Eq40}
\end{equation}%
and write%
\begin{equation}
dY_{\upsilon }=dW_{\upsilon },\ \ \ Y_{0}=0,\ \ \ Y_{\upsilon }\geq \xi
-N_{\upsilon },  \label{Eq41}
\end{equation}%
where 
\begin{equation}
N_{\upsilon }=M_{t\left( \upsilon \right) }.  \label{Eq41a}
\end{equation}

Thus, we have one of the two venues to explore: (A) studying the processes $%
X_{t}$ or $X_{\upsilon }$ given by Eq. (\ref{Eq36}) and Eq. (\ref{Eq37}),
respectively; (B) dealing with the processes $Y_{t}$ or $Y_{\upsilon }$
given by Eq. (\ref{Eq39}) and Eq. (\ref{Eq41}). In case (A) there is
non-zero time-dependent drift and flat boundary; in case (B) there is no
drift but the boundary is time-dependent.

Given a variance trajectory, we need to discuss how to calculate $\mu ,\nu
,\lambda ,M$, and $N$. Let $\left\{ v_{k}|k=0,1,...,K\right\} $, $v_{0}=v$,
be a particular path generated via discretization of Eqs (\ref{Eq13}) with
homogeneous time-step $\Delta t=T/K$. This equation can be discretized in
various ways, for example, via the Euler-Maryama scheme or the Milstein
scheme. For special cases such as the Feller process corresponding to the
Heston model, there are clever schemes tailored to the specific process at
hand. However, we are not pursuing them here since it is unnecessary to
achieve our objective. We have%
\begin{equation}
\begin{array}{c}
v_{0}=v,\ \ \ v_{k}=v_{k-1}+\kappa \left( \theta -v_{k-1}\right) \Delta
t+\varepsilon \sqrt{\Delta t}\eta , \\ 
I_{0}=0,\ \ \ I_{k}=I_{k-1}+\frac{\Delta t}{2}\left( v_{k}+v_{k-1}\right) ,%
\end{array}
\label{Eq42}
\end{equation}%
\begin{equation}
\begin{array}{c}
\mu_{k}=r-\frac{\rho \kappa \theta }{\varepsilon }-\left( \frac{1}{2}-\frac{%
\rho \kappa }{\varepsilon }\right) v_{k}+\frac{\rho }{\varepsilon }\frac{%
\left( \left( v_{k}-v_{k-1}\right) \right) }{\Delta t}, \\ 
\nu _{k}=\sqrt{1-\rho ^{2}}\sqrt{v_{k}}, \\ 
M_{0}=0,\ \ \ M_{k}=\frac{kT}{K}\left( r-\frac{\rho \kappa \theta }{%
\varepsilon }\right) -\left( \frac{1}{2}-\frac{\rho \kappa }{\varepsilon }%
\right) I_{k}+\frac{\rho }{\varepsilon }\left( v_{k}-v\right) .%
\end{array}
\label{Eq43}
\end{equation}%
It is clear that%
\begin{equation}
\upsilon _{k}=\left( 1-\rho ^{2}\right) I_{k},\ \ \ \Upsilon =\left( 1-\rho
^{2}\right) I_{K},  \label{Eq44}
\end{equation}%
where $\left\{ \upsilon _{k}|k=0,1,...,K\right\} $ is an inhomogeneous
partition of the interval $\left[ 0,\Upsilon \right] $. For our purposes, we
treat the sequences $\left\{ \lambda _{k}=\mu _{k}/\nu
_{k}|k=0,1,...,K\right\} $ and \newline
$\left\{ M_{k}|k=0,1,...,K\right\} $ as functions of $\upsilon _{k}$, which
is possible because $\upsilon _{k}$ is a monotonically increasing sequence,
and interpret them accordingly.

\begin{figure}[tbp]
\begin{center}
\subfloat[]{\includegraphics[width=0.75\textwidth]
{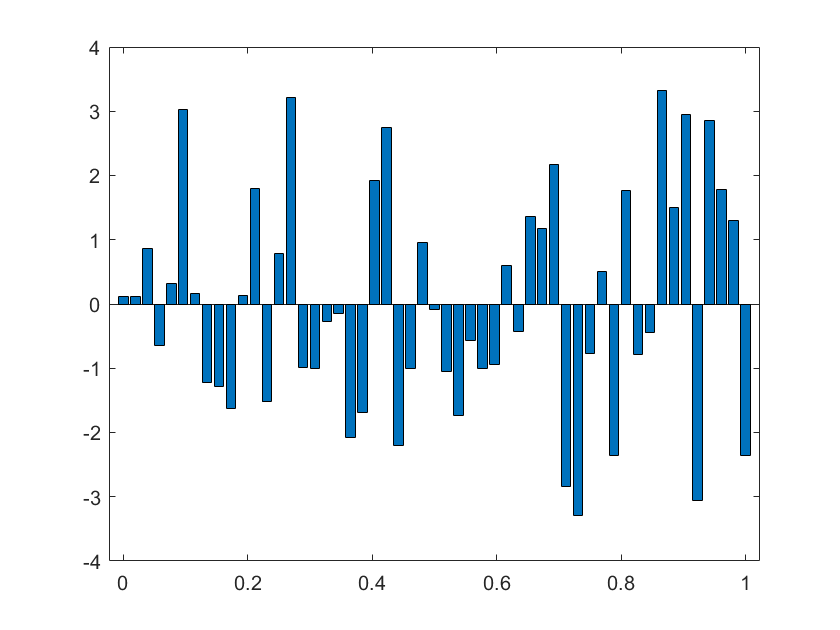}} \\%
[0pt]
\subfloat[]{\includegraphics[width=0.75\textwidth]
{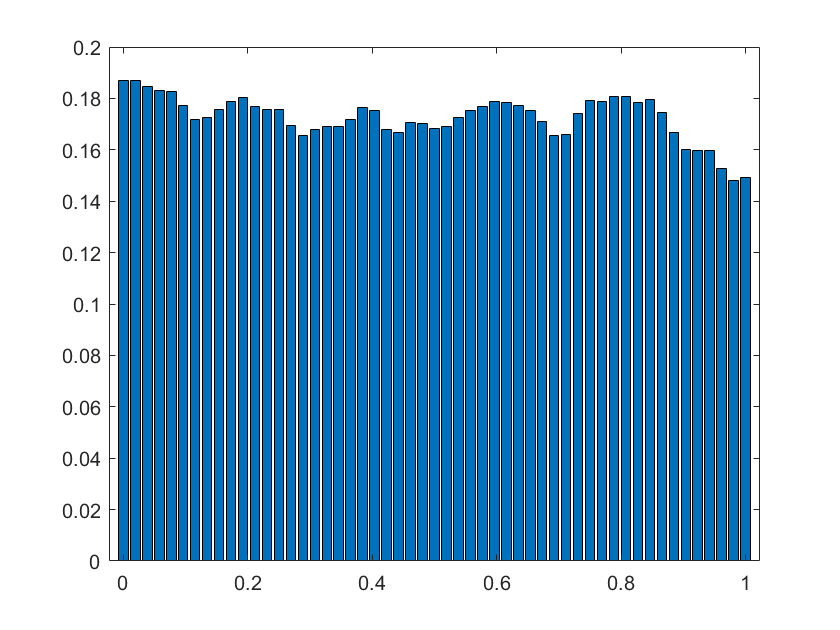}}\\[%
0pt]
\end{center}
\par
\vspace{-10pt}
\caption{Time-dependent coefficients (a) advection, (b) diffusion. The
corresponding parameters are the same as in Figure \protect\ref{Fig1}. Here $%
N_{t}=52$ - one step per week.}
\label{Fig2}
\end{figure}

\begin{figure}[tbp]
\begin{center}
\subfloat[]{\includegraphics[width=0.75\textwidth]
{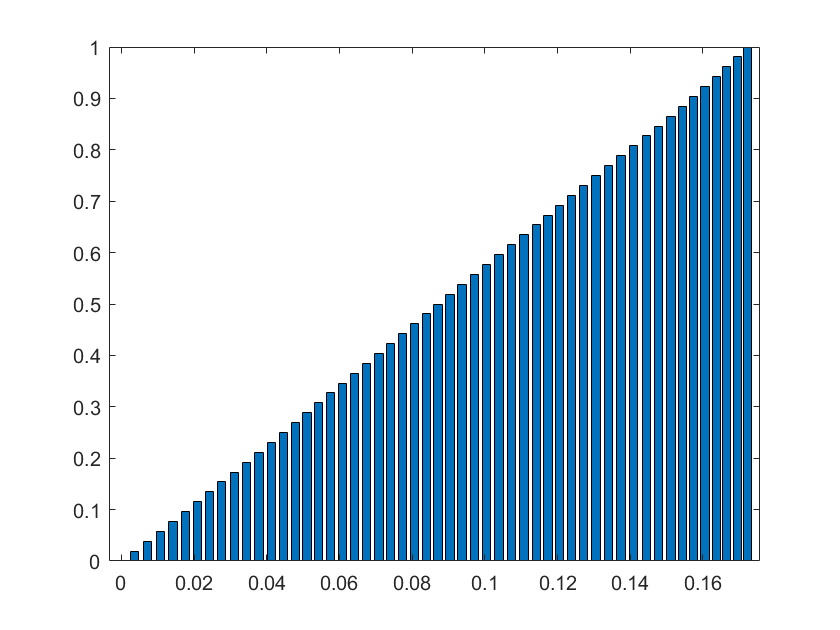}} \\%
[0pt]
\subfloat[]{\includegraphics[width=0.75\textwidth]
{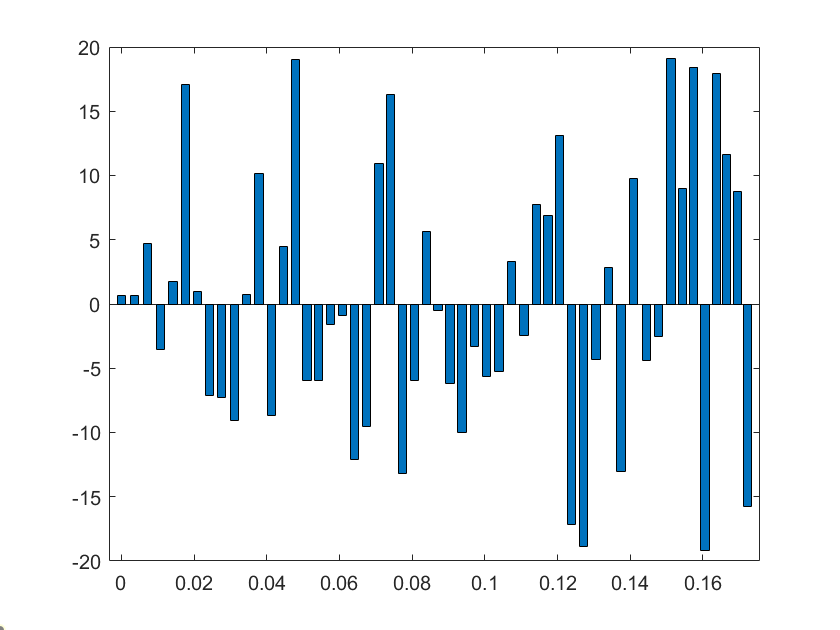}}\\[%
0pt]
\end{center}
\par
\vspace{-10pt} .
\caption{(a) $t\left( \Upsilon \right) $, (b) $\protect\lambda \left(
\Upsilon \right) $}
\label{Fig3}
\end{figure}

We illustrate the corresponding functions in Figures \ref{Fig2}, \ref{Fig3}. 
We emphasize that $\mu $ and $\lambda $ are very irregular, since they
depend on the white noise process $dW_{t}/dt$, so that we have to deal with
random terms of order $\Delta t^{-1/2}$. At the same time, the moving
boundary $M_{t}$ is much more regular, and depends on random terms of order $%
\Delta t^{1/2}$.

\subsection{Solving IBVPs via Crank-Nicolson method}

Let us describe how to price a conditional barrier option via the FDM; see 
\cite{Loeper2009}.\footnote{%
Surprisingly, \cite{Loeper2009} do not comment on the irregular nature of
the drift coefficient.} Specifically, we want to solve the following problem
on the semi-axis $\left( \xi ,\infty \right) $:%
\begin{equation}
\begin{array}{c}
P_{t}\left( t,X\right) +\mu \left( t\right) P_{X}\left( t,X\right) +\frac{1}{%
2}\nu \left( t\right) P_{XX}\left( t,X\right) -\varkappa \left( t\right)
P\left( t,X\right) =0, \\ 
P\left( T,X\right) =\Pi \left( X\right) ,\ \ \ P(t,\xi )=0.%
\end{array}
\label{Eq45}
\end{equation}%
We introduce $\bar{P}$, such that%
\begin{equation}
P\left( t,X\right) =e^{-\int_{t}^{T}\varkappa \left( t^{\prime }\right)
dt^{\prime }}\bar{P}\left( t,X\right) ,  \label{Eq45a}
\end{equation}%
and get the following problem for $\bar{P}\left( t,X\right) $:%
\begin{equation}
\begin{array}{c}
\bar{P}_{t}\left( t,X\right) +\mu \left( t\right) \bar{P}_{X}\left(
t,X\right) +\frac{1}{2}\nu \left( t\right) \bar{P}_{XX}\left( t,X\right) =0,
\\ 
\bar{P}\left( T,X\right) =\Pi \left( X\right) ,\ \ \ \bar{P}(t,\xi )=0.%
\end{array}
\label{Eq45b}
\end{equation}%
In the context we are interested in, $\varkappa \left( t\right) =r$, so that 
$P=\exp \left( r\left( t-T\right) \right) \bar{P}$.

We choose a uniform spatial grid $\left\{ \left. x_{m}=\xi +m\left( U-\xi
\right) /M\right\vert m=0,1,...,M\right\} $, where $U$ is a sufficiently
remote upper boundary such that $0$ belongs to the spatial grid, $x_{o}=0$,
and a temporal grid $\left\{ \left. t^{n}\right\vert
t^{0}<t^{1}<...<t^{N}=T\right\} $, which is not necessarily uniform. Then,
we apply the usual Crank-Nicolson method for the advection-diffusion
diffusion and get the following system of matrix equations for $\bar{P}%
_{m}^{n}$:%
\begin{equation}
\begin{array}{c}
\bar{P}_{m}^{n+1}-\bar{P}_{m}^{n}+\alpha ^{n+1/2}\left( \bar{P}_{m+1}^{n+1}-%
\bar{P}_{m-1}^{n+1}+\bar{P}_{m+1}^{n}-\bar{P}_{m-1}^{n}\right) \\ 
+\beta ^{n+1/2}\left( \bar{P}_{m+1}^{n+1}-2\bar{P}_{m}^{n+1}+\bar{P}%
_{m-1}^{n+1}+\bar{P}_{m+1}^{n}-2\bar{P}_{m}^{n}+\bar{P}_{m-1}^{n}\right) =0.%
\end{array}
\label{Eq46}
\end{equation}%
where%
\begin{equation}
\begin{array}{c}
\alpha ^{n+1/2}=\frac{\mu ^{n+1/2}\Delta t^{n+1/2}}{4\Delta x},\ \ \ \beta
^{n+1/2}=\frac{\nu ^{n+1/2}\Delta t^{n+1/2}}{4\Delta x^{2}}.%
\end{array}
\label{Eq47}
\end{equation}%
We emphasize that given the extreme irregularity of $\mu $, using more
complicated approaches for treading the drift term is not warranted.

We can write the system of equations in matrix form.%
\begin{equation}
\begin{array}{c}
\mathbb{A}_{\left( \alpha ,\beta \right) }^{n+1/2}\bar{P}^{n}=\mathbb{A}%
_{\left( -\alpha ,-\beta \right) }^{n+1/2}\bar{P}^{n+1}, \\ 
\bar{P}_{m}^{N}=\Pi \left( x_{m}\right) ,%
\end{array}
\label{Eq48}
\end{equation}%
where%
\begin{equation}
\begin{array}{c}
\mathbb{A}_{\left( \alpha ,\beta ,\gamma \right) }=\left( 
\begin{array}{cccccc}
1 & 0 & 0 & 0 & 0 & 0 \\ 
\alpha -\beta & 1+2\beta & -\alpha -\beta & 0 & 0 & 0 \\ 
. & . & . & . & . & . \\ 
. & . & . & . & . & . \\ 
0 & 0 & 0 & \alpha -\beta & 1+2\beta & -\alpha -\beta \\ 
0 & 0 & \beta & -\alpha -4\beta & 4\alpha +5\beta & 1-3\alpha -2\beta%
\end{array}%
\right) .%
\end{array}
\label{Eq49}
\end{equation}%
For brevity, the superscripts are omitted. At infinity we apply one-sided
derivatives and use the PDE \emph{itself} to formulate the boundary
condition.

Alternatively, we can rescale time, reduce the problem (\ref{Eq29}) to the
form:%
\begin{equation}
\begin{array}{c}
\bar{P}_{\upsilon }+\lambda \left( \upsilon \right) \bar{P}_{x}+\frac{1}{2}%
\bar{P}_{xx}=0, \\ 
\bar{P}\left( \Upsilon ,x\right) =\Pi \left( x\right) ,\ \ \ \bar{P}%
(\upsilon ,\xi )=0,%
\end{array}
\label{Eq51}
\end{equation}%
and apply the Crank-Nicolson method to problem (\ref{Eq51}).

Of course, we can attack the pricing problem by solving the corresponding
Fokker-Planck equation for Green's function $G$:%
\begin{equation}
\begin{array}{c}
G_{t}(t,X)+\mu \left( t\right) G_{X}(t,X)-\frac{1}{2}\nu \left( t\right)
G_{XX}(t,X)+\varkappa \left( t\right) G(t,X)=0, \\ 
G\left( 0,X\right) =\delta \left( X\right) ,\ \ \ G(t,\xi )=0,%
\end{array}
\label{Eq52}
\end{equation}%
and write%
\begin{equation}
\begin{array}{c}
P\left( 0,0\right) =\int\limits_{0}^{\infty }G\left( 0,0;T,x\right) \Pi
\left( x\right) dx.%
\end{array}
\label{Eq53}
\end{equation}%
We introduce $\bar{G}$, such that%
\begin{equation}
G\left( t,X\right) =e^{-\int_{0}^{t}\varkappa \left( t^{\prime }\right)
dt^{\prime }}\bar{G}\left( t,X\right) .  \label{Eq53a}
\end{equation}%
Accordingly,%
\begin{equation}
\begin{array}{c}
\bar{G}_{t}(t,X)+\mu \left( t\right) \bar{G}_{X}(t,X)-\frac{1}{2}\nu \left(
t\right) \bar{G}_{XX}(t,X)=0, \\ 
\bar{G}\left( 0,X\right) =\delta \left( X\right) ,\ \ \ \bar{G}(t,\xi )=0,%
\end{array}
\label{Eq53b}
\end{equation}%
\begin{equation}
\begin{array}{c}
P\left( 0,0\right) =e^{-\int_{0}^{T}\varkappa \left( t^{\prime }\right)
dt^{\prime }}\int\limits_{0}^{\infty }\bar{G}\left( 0,0;T,x\right) \Pi
\left( x\right) dx.%
\end{array}%
\end{equation}%
As before, we apply the Crank-Nicolson method for the advection-diffusion
and get the following system of matrix equations for $\bar{G}_{m}^{n}$:%
\begin{equation}
\begin{array}{c}
\bar{G}_{m}^{n+1}-\bar{G}_{m}^{n}+\alpha ^{n+1/2}\left( \bar{G}_{m+1}^{n+1}-%
\bar{G}_{m-1}^{n+1}+\bar{G}_{m+1}^{n}-\bar{G}_{m-1}^{n}\right) \\ 
-\beta ^{n+1/2}\left( \bar{G}_{m+1}^{n+1}-2\bar{G}_{m}^{n+1}+\bar{G}%
_{m-1}^{n+1}+\bar{G}_{m+1}^{n}-2\bar{G}_{m}^{n}+\bar{G}_{m-1}^{n}\right) =0.%
\end{array}
\label{Eq54}
\end{equation}%
In the matrix form we get%
\begin{equation}
\begin{array}{c}
\mathbb{A}_{\left( \alpha ,\beta \right) }^{n+1/2}\bar{G}^{n+1}=\mathbb{A}%
_{\left( -\alpha ,-\beta \right) }^{n+1/2}\bar{G}^{n}, \\ 
\bar{G}^{0}=\delta _{o,m}.%
\end{array}
\label{Eq55}
\end{equation}%
Once $\bar{G}^{N}$ is found, we can represent $P_{k}^{0}$ as%
\begin{equation}
\begin{array}{c}
P_{k}^{0}=e^{-rT}\Delta x\sum\limits_{m=1}^{M}\bar{G}_{m}^{N}\Pi _{m}.%
\end{array}
\label{Eq56}
\end{equation}

\subsection{Solving IBVPs via MHP\label{Sec42}}

\subsubsection{Analytic results}

\begin{figure}[tbp]
\begin{center}
{%
\includegraphics[width=0.8\textwidth]
{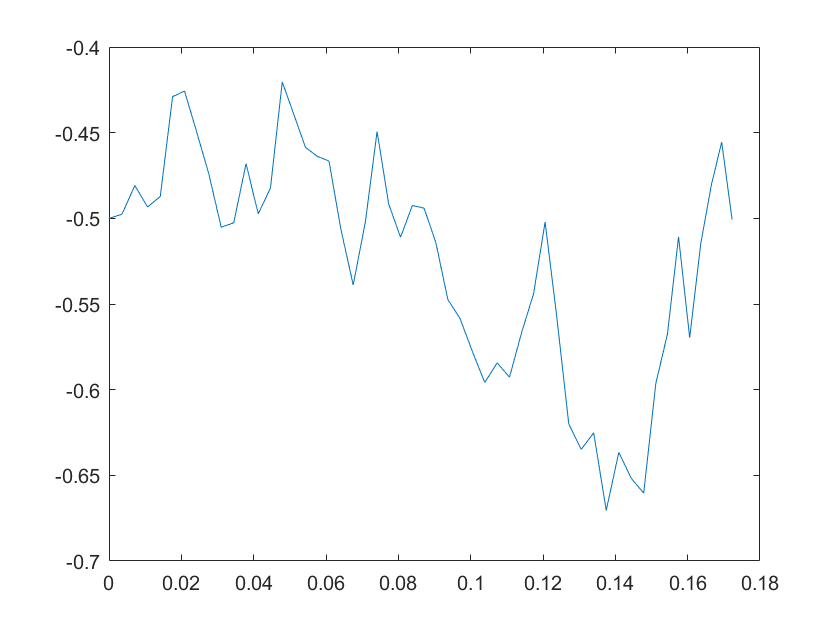}}
\end{center}
\par
\vspace{-10pt}
\caption{A typical lower boundary $\protect\xi -N_{\protect\upsilon }$ as a
function of $\protect\upsilon $.}
\label{Fig4}
\end{figure}

We wish to find Green's function for process (\ref{Eq41}), or, equivalently,
to solve the following IBVP in a bounded domain with a moving boundary:

\begin{equation}
\begin{array}{c}
\frac{\partial }{\partial \upsilon }\bar{G}\left( \upsilon ,Y\right) =\frac{1%
}{2}\frac{\partial ^{2}}{\partial Y^{2}}\bar{G}\left( \upsilon ,Y\right) ,\
\  \\ 
\bar{G}\left( 0,Y\right) =\delta \left( Y\right) ,\ \ \ \bar{G}\left(
\upsilon ,\xi -N_{\upsilon }\right) =0,\ \ \ \bar{G}\left( \upsilon
,Y\rightarrow \infty \right) \rightarrow 0, \\ 
\ \xi -N_{\upsilon }\leq Y<\infty ,\ \ \ \ \ 0\leq \upsilon \leq \Upsilon .%
\end{array}
\label{Eq57}
\end{equation}%
For a representative Monte Carlo path, the corresponding boundary is shown
in Figure \ref{Fig4}. 

We split $\bar{G}$, and represent it in the form:%
\begin{equation}
\bar{G}\left( \upsilon ,Y\right) =H\left( \upsilon ,Y\right) -F\left(
\upsilon ,Y\right) ,  \label{Eq58}
\end{equation}%
where $H\left( \upsilon ,Y\right) $ is the standard heat kernel,%
\begin{equation}
H\left( \upsilon ,Y\right) =\frac{\exp \left( -\frac{Y^{2}}{2\upsilon }%
\right) }{\sqrt{2\pi \upsilon }},  \label{Eq59}
\end{equation}%
and $F\left( \upsilon ,Y\right) $ solves the following problem:%
\begin{equation}
\begin{array}{c}
\frac{\partial }{\partial \upsilon }F\left( \upsilon ,Y\right) =\frac{1}{2}%
\frac{\partial ^{2}}{\partial Y^{2}}F\left( \upsilon ,Y\right) ,\ \ \ \xi
-N_{\upsilon }\leq Y<\infty , \\ 
F\left( 0,Y\right) =0,\ \ \ F\left( \upsilon ,\xi -N_{\upsilon }\right)
=f\left( \upsilon \right) ,\ \ \ F\left( \upsilon ,Y\rightarrow \infty
\right) \rightarrow 0,%
\end{array}
\label{Eq60}
\end{equation}%
where%
\begin{equation}
f\left( \upsilon \right) =H\left( \upsilon ,\xi -N_{\upsilon }\right) .
\label{Eq61}
\end{equation}%
The MHP allows one to represent $F\left( \upsilon ,Y\right) $ in the form%
\begin{equation}
F\left( \upsilon ,Y\right) =\int_{0}^{\upsilon }\frac{\left( Y-\xi
+N_{\upsilon ^{\prime }}\right) \exp \left( -\frac{\left( Y-\xi +N_{\upsilon
^{\prime }}\right) ^{2}}{2\left( \upsilon -\upsilon ^{\prime }\right) }%
\right) }{\sqrt{2\pi \left( \upsilon -\upsilon ^{\prime }\right) ^{3}}}\phi
\left( \upsilon ^{\prime }\right) d\upsilon ^{\prime },  \label{Eq62}
\end{equation}%
where $\phi \left( \upsilon \right) $ solves the Volterra equation of the
second kind:%
\begin{equation}
\phi \left( \upsilon \right) +\int_{0}^{\upsilon }\frac{\Theta \left(
\upsilon ,\upsilon ^{\prime }\right) \Xi \left( \upsilon ,\upsilon ^{\prime
}\right) }{\sqrt{2\pi \left( \upsilon -\upsilon ^{\prime }\right) }}\phi
\left( \upsilon ^{\prime }\right) d\upsilon ^{\prime }=f\left( \upsilon
\right) ,  \label{Eq63}
\end{equation}%
and%
\begin{equation}
\begin{array}{c}
\ \Theta \left( \upsilon ,\upsilon ^{\prime }\right) =-\frac{N_{\upsilon
}-N_{\upsilon ^{\prime }}}{\left( \upsilon -\upsilon ^{\prime }\right) },\ \
\ \Xi \left( \upsilon ,\upsilon ^{\prime }\right) =\exp \left( -\frac{\left(
\upsilon -\upsilon ^{\prime }\right) \Theta ^{2}\left( \upsilon ,\upsilon
^{\prime }\right) }{2}\right) , \\ 
\Theta \left( \upsilon ,\upsilon \right) =-\frac{dN_{\upsilon }}{d\upsilon }%
,\ \ \ \Xi ^{>}\left( \upsilon ,\upsilon \right) =1.%
\end{array}
\label{Eq64}
\end{equation}%
Assuming that $\phi \left( \upsilon \right) $ is known, we can represent $%
\bar{G}\left( \Upsilon ,Y\right) $ as follows:%
\begin{equation}
\bar{G}\left( \Upsilon ,Y\right) =H\left( \Upsilon ,Y\right) -F\left(
\Upsilon ,Y\right) ,  \label{Eq65}
\end{equation}%
where $F\left( \Upsilon ,Y\right) $ is given by Eq. (\ref{Eq62}). Finally,
returning back to the original variables, we get%
\begin{equation}
\bar{G}\left( T,X\right) =H\left( \Upsilon \left( T\right) ,X-M_{T}\right)
-F\left( \Upsilon \left( T\right) ,X-M_{T}\right) .  \label{Eq67}
\end{equation}

When $Z=X-\xi \rightarrow 0$, the corresponding integral has to be dealt
with carefully due to a singularity at $\upsilon ^{\prime }=\upsilon $. We
have

\begin{equation}
\begin{array}{c}
F\left( \upsilon ,X-N_{\upsilon }\right) =\int_{0}^{\upsilon }\frac{\left(
Z-N_{\upsilon }+N_{\upsilon ^{\prime }}\right) \exp \left( -\frac{\left(
Z-N_{\upsilon }+N_{\upsilon ^{\prime }}\right) ^{2}}{2\left( \upsilon
-\upsilon ^{\prime }\right) }\right) }{\sqrt{2\pi \left( \upsilon -\upsilon
^{\prime }\right) ^{3}}}\phi \left( \upsilon ^{\prime }\right) d\upsilon
^{\prime } \\ 
=e^{ZN^{\prime }\left( \upsilon \right) }\phi \left( \upsilon \right)
\int_{0}^{\upsilon }\frac{Z\exp \left( -\frac{Z^{2}}{2\left( \upsilon
-\upsilon ^{\prime }\right) }\right) }{\sqrt{2\pi \left( \upsilon -\upsilon
^{\prime }\right) ^{3}}}d\upsilon ^{\prime }+\int_{0}^{\upsilon }\frac{%
I^{\left( 1\right) }\left( \upsilon ,\upsilon ^{\prime }\right) }{\sqrt{%
\left( \upsilon -\upsilon ^{\prime }\right) }}d\upsilon ^{\prime
}+\int_{0}^{\upsilon }\frac{I^{\left( 2\right) }\left( \upsilon ,\upsilon
^{\prime }\right) }{\sqrt{\left( \upsilon -\upsilon ^{\prime }\right) }}%
d\upsilon ^{\prime } \\ 
=2e^{ZN^{\prime }\left( \upsilon \right) }\phi \left( \upsilon \right) 
\mathfrak{N}\left( -\frac{Z}{\sqrt{\upsilon }}\right) +\int_{0}^{\upsilon }%
\frac{I^{\left( 1\right) }\left( \upsilon ,\upsilon ^{\prime }\right) }{%
\sqrt{\left( \upsilon -\upsilon ^{\prime }\right) }}d\upsilon ^{\prime
}+\int_{0}^{\upsilon }\frac{I^{\left( 2\right) }\left( \upsilon ,\upsilon
^{\prime }\right) }{\sqrt{\left( \upsilon -\upsilon ^{\prime }\right) }}%
d\upsilon ^{\prime },%
\end{array}
\label{Eq66}
\end{equation}%
where%
\begin{equation}
\begin{array}{c}
I^{\left( 1\right) }\left( \upsilon ,\upsilon ^{\prime }\right) =\frac{Z\exp
\left( -\frac{Z^{2}}{2\left( \upsilon -\upsilon ^{\prime }\right) }\right)
\left( \exp \left( -Z\Theta \left( \upsilon ,\upsilon ^{\prime }\right)
\right) \Xi \left( \upsilon ,\upsilon ^{\prime }\right) \phi \left( \upsilon
^{\prime }\right) -e^{ZN^{\prime }\left( \upsilon \right) }\phi \left(
\upsilon \right) \right) }{\sqrt{2\pi }\left( \upsilon -\upsilon ^{\prime
}\right) }, \\ 
I^{\left( 2\right) }\left( \upsilon ,\upsilon ^{\prime }\right) =\frac{%
\Theta \left( \upsilon ,\upsilon ^{\prime }\right) \exp \left( -\frac{\left(
Z-N_{\upsilon }+N_{\upsilon ^{\prime }}\right) ^{2}}{2\left( \upsilon
-\upsilon ^{\prime }\right) }\right) \phi \left( \upsilon ^{\prime }\right) 
}{\sqrt{2\pi }}.%
\end{array}
\label{Eq66a}
\end{equation}%
It is clear that integrals in Eq. (\ref{Eq66}) have weak singularities and
hence are easy to handle.

\subsubsection{Numerics}

There are numerous well-known approaches to solving Volterra equations; see, 
\cite{Linz1985}, among many others. We choose the most straightforward
approach and show how to solve the following archetypal Volterra equation
with weak singularity numerically: 
\begin{equation}
\phi (\upsilon )+\int_{0}^{\upsilon }\frac{K(\upsilon ,\upsilon ^{\prime })}{%
\sqrt{\upsilon -\upsilon ^{\prime }}}\phi (\upsilon ^{\prime })\,d\upsilon
^{\prime }=f(\upsilon ),  \label{Eq68}
\end{equation}%
where $K(\upsilon ,\upsilon ^{\prime })$ is a non-singular kernel. We write

\begin{equation}
\int_{0}^{\upsilon }\frac{K(\upsilon ,\upsilon ^{\prime })\phi \left(
\upsilon ^{\prime }\right) }{\sqrt{\upsilon -\upsilon ^{\prime }}}d\upsilon
^{\prime }=-2\int_{0}^{\upsilon }K(\upsilon ,\upsilon ^{\prime })\phi \left(
\upsilon ^{\prime }\right) \,d\sqrt{\upsilon -\upsilon ^{\prime }}.
\label{Eq69}
\end{equation}%
We map this equation to a grid $0=\upsilon _{0}<\upsilon _{1}<\ldots
<\upsilon _{N}=\upsilon $. We introduce the following notation: 
\begin{equation}
\begin{array}{c}
f_{k}=f(\upsilon _{k}),\ \ \ \phi _{k}=\phi \left( \upsilon _{k}\right) ,\ \
\ K_{k,l}=K(\upsilon _{k},\upsilon _{l}), \\ 
\Delta _{k,l}=\upsilon _{k}-\upsilon _{l},\ \ \ \Pi _{k,l}=\frac{\Delta
_{l,l-1}}{\left( \sqrt{\Delta _{k,l-1}}+\sqrt{\Delta _{k,l}}\right) }.%
\end{array}
\label{Eq70}
\end{equation}%
Then, Eq. (\ref{Eq68}) can be approximated by the trapezoidal rule as

\begin{equation}
\phi _{k}+\sum_{l=1}^{k}\Pi _{k,l}\left( K_{k,l}\phi _{l}+K_{k,l-1}\phi
_{l-1}\right) =f_{k},  \label{Eq71}
\end{equation}%
so that

\begin{equation}
\phi _{k}=\frac{\left( f_{k}-\sqrt{\Delta _{k,k-1}}K_{k,k-1}\phi
_{k-1}-\sum_{l=1}^{k-1}\Pi _{k,l}\left( K_{k,l}\phi _{l}+K_{k,l-1}\phi
_{l-1}\right) \right) }{\left( 1+\sqrt{\Delta _{k,k-1}}K_{k,k}\right) }.
\label{Eq72}
\end{equation}%
Thus, $\phi _{k}$ can be found by induction starting with $\phi _{0}=f_{0}$.

After $\phi _{k}$ are determined, $F_{k}$ can be written in the form%
\begin{equation}
F_{k}=2e^{ZN_{k}^{\prime }}\phi _{k}\mathfrak{N}\left( -\frac{Z}{\sqrt{%
\upsilon _{k}}}\right) +\sum_{l=1}^{k}\Pi _{k,l}\left( I_{k,l}^{\left(
1\right) }+I_{k,l-1}^{\left( 1\right) }\right) +\sum_{l=1}^{k}\Pi
_{k,l}\left( I_{k,l}^{\left( 2\right) }+I_{k,l-1}^{\left( 2\right) }\right) ,
\label{Eq72a}
\end{equation}

\subsubsection{Example}

\begin{figure}[tbp]
\begin{center}
{%
\includegraphics[width=0.8\textwidth]
{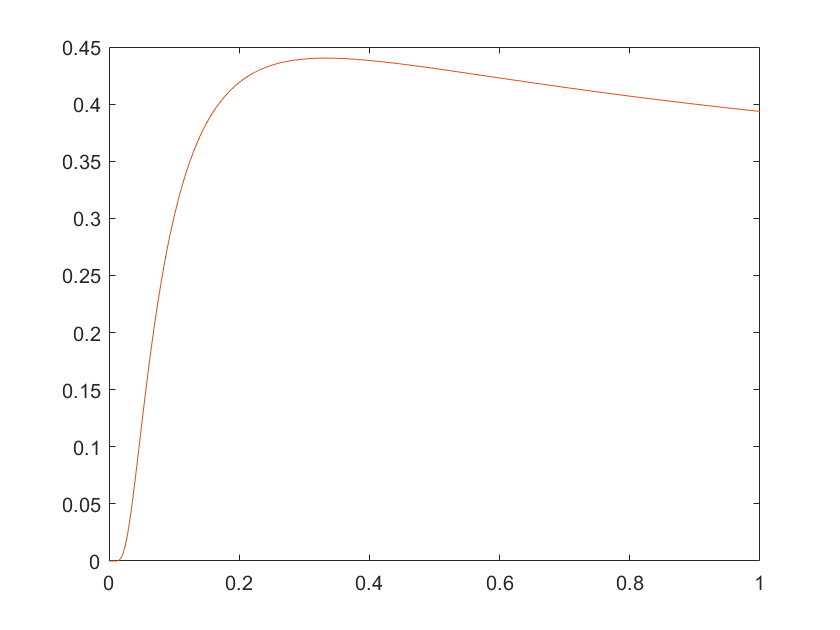}}
\end{center}
\par
\vspace{-10pt}
\caption{$\protect\phi \left( \protect\upsilon \right) $ computed
analytically and by solving the Volterra equation. The difference between
the corresponding functions is less that $10^{(}-4)$. The parameters are $%
\protect\xi =-0.5$, $\protect\lambda =0.5$.}
\label{Fig5}
\end{figure}

Let us consider the special case of constant drift $\lambda $; the
corresponding boundary is linear, $\xi -\lambda \upsilon $, where $\upsilon
=t$. Then Eq. (\ref{Eq63}) becomes%
\begin{equation}
\phi \left( \upsilon \right) -\lambda \int_{0}^{\upsilon }\frac{\exp \left(
^{-}\frac{\lambda ^{2}\left( \upsilon -\upsilon ^{\prime }\right) }{2}%
\right) }{\sqrt{2\pi \left( \upsilon -\upsilon ^{\prime }\right) }}\phi
\left( \upsilon ^{\prime }\right) d\upsilon ^{\prime }=\frac{e^{-\frac{%
\left( \xi -\lambda \upsilon \right) ^{2}}{2\upsilon }}}{\sqrt{2\pi \upsilon 
}}.  \label{Eq74}
\end{equation}%
\cite{Lipton2020} show that%
\begin{equation}
\phi \left( \upsilon \right) =\frac{e^{-\frac{\left( \xi -\lambda \upsilon
\right) ^{2}}{2\upsilon }}}{\sqrt{2\pi \upsilon }}+\lambda e^{2\xi \lambda }%
\mathfrak{N}\left( \frac{\xi +\lambda \upsilon }{\sqrt{\upsilon }}\right) ,
\label{Eq75}
\end{equation}%
\begin{equation}
F\left( \upsilon ,Y\right) =\frac{\exp \left( 2\xi \lambda -\frac{\left(
Y-2\xi \right) ^{2}}{2\upsilon }\right) }{\sqrt{2\pi \upsilon }}.
\label{Eq75a}
\end{equation}%
Accordingly,%
\begin{equation}
\bar{G}\left( T,X\right) =H\left( T,X-\lambda T\right) -e^{2\xi \lambda
}H\left( T,X-\lambda T-2\xi \right) .  \label{Eq76}
\end{equation}%
It is easy to see that $\bar{G}\left( T,\xi \right) =0$, as it should.
Figure \ref{Fig5} illustrates our findings.

\subsubsection{Semi-analytical solution of pricing problems}

\begin{figure}[tbp]
\begin{center}
\includegraphics[width=0.8\textwidth]
{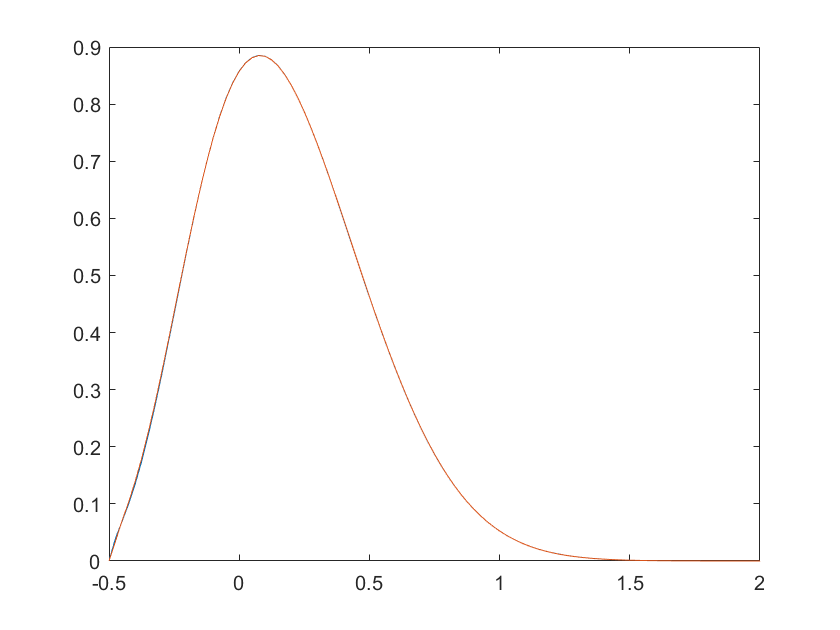}
\end{center}
\par
\vspace{-10pt}
\caption{$G\left( T,x\right) $ computed via the FDM and the MHP. The
absolute difference between the corresponding functions is less that $%
10^{-3} $. The parameters are the same as in Figure \protect\ref{Fig1}. The
log-barrier is $\protect\xi =-0.5.$}
\label{Fig6}
\end{figure}

Green's function is given by Eq. (\ref{Eq67}). In Figure \ref{Fig6} we
compare Green's functions obtained via the FDM and the MHP. The figure shows
that these functions are reassuringly close. The MHP is clearly preferred
since it is much faster. 

Once $\bar{G}\left( T,X\right) $ is found all barrier problems can be solved
analytically.

For instance, we can calculate the $P\left( 0,T;\xi \right) $ of the
no-touch option for the barrier level $\xi <0$: 
\begin{equation}
\begin{array}{c}
P\left( 0,T;\xi \right) =e^{-rT}\int_{\xi }^{\infty }\left( H\left( \Upsilon
,X^{\prime }-N_{\Upsilon }\right) -F\left( \Upsilon ,X^{\prime }-N_{\Upsilon
}\right) \right) dX^{\prime } \\ 
=e^{-rT}\left( \mathfrak{N}\left( \frac{-\left( \xi -N_{\Upsilon }\right) }{%
\sqrt{\Upsilon }}\right) -\int_{0}^{\Upsilon }\frac{\exp \left( -\frac{%
\left( \xi -N_{\Upsilon }+N_{\upsilon ^{\prime }}\right) ^{2}}{2\left(
\Upsilon -\upsilon ^{\prime }\right) }\right) }{\sqrt{2\pi \left( \Upsilon
-\upsilon ^{\prime }\right) }}\phi \left( \upsilon ^{\prime }\right)
d\upsilon ^{\prime }\right) .%
\end{array}
\label{Eq77}
\end{equation}%
The relative price $C\left( 0,T,K;\xi \right) /e^{-rT}S_{0}$ of the barrier
call struck at $K=S_{0}e^{k}$, where $k\geq \xi $, has the form:

\begin{equation}
\begin{array}{c}
\frac{C\left( 0,S_{0},T,K;\xi \right) }{e^{-rT}S_{0}} \\ 
=\int_{k}^{\infty }\left( H\left( \Upsilon ,X^{\prime }-N_{\Upsilon }\right)
-F\left( \Upsilon ,X^{\prime }-N_{\Upsilon }\right) \right) \left(
e^{X^{\prime }}-e^{k}\right) dX^{\prime } \\ 
=e^{N_{\Upsilon }}\int_{k-N_{\Upsilon }}^{\infty }\frac{e^{-\frac{\eta ^{2}}{%
2\Upsilon }+\eta }}{\sqrt{2\pi \Upsilon }}d\eta -e^{k}\int_{k-N_{\Upsilon
}}^{\infty }\frac{e^{-\frac{\eta ^{2}}{2\Upsilon }}}{\sqrt{2\pi \Upsilon }}%
d\eta \\ 
-e^{N_{\Upsilon }}\int_{0}^{\Upsilon }\int_{k-N_{\Upsilon }}^{\infty }\frac{%
\left( \eta -\xi +N_{\upsilon ^{\prime }}\right) \exp \left( -\frac{\left(
\eta -\xi +N_{\upsilon ^{\prime }}\right) ^{2}}{2\left( \Upsilon -\upsilon
^{\prime }\right) }+\eta \right) }{\sqrt{2\pi \left( \Upsilon -\upsilon
^{\prime }\right) ^{3}}}d\eta \phi \left( \upsilon ^{\prime }\right)
d\upsilon ^{\prime } \\ 
+e^{k}\int_{0}^{\Upsilon }\int_{k-N_{\Upsilon }}^{\infty }\frac{\left( \eta
-\xi +N_{\upsilon ^{\prime }}\right) \exp \left( -\frac{\left( \eta -\xi
+N_{\upsilon ^{\prime }}\right) ^{2}}{2\left( \Upsilon -\upsilon ^{\prime
}\right) }\right) }{\sqrt{2\pi \left( \Upsilon -\upsilon ^{\prime }\right)
^{3}}}d\eta \phi \left( \upsilon ^{\prime }\right) d\upsilon ^{\prime } \\ 
=e^{N_{\Upsilon }+\frac{\Upsilon }{2}}\mathfrak{N}\left( \frac{N_{\Upsilon
}+\Upsilon }{\sqrt{\Upsilon }}\right) -e^{k}\mathfrak{N}\left( \frac{%
N_{\Upsilon }}{\sqrt{\Upsilon }}\right) \\ 
-e^{k}\int_{0}^{\Upsilon }\frac{\exp \left( -\frac{\left( k-N_{\Upsilon
}-\xi +N_{\upsilon ^{\prime }}\right) ^{2}}{2\left( \Upsilon -\upsilon
^{\prime }\right) }\right) }{\sqrt{2\pi \left( \Upsilon -\upsilon ^{\prime
}\right) }}\phi \left( \upsilon ^{\prime }\right) d\upsilon ^{\prime } \\ 
-\int_{0}^{\Upsilon }e^{N_{\Upsilon }+\xi -N_{\upsilon ^{\prime }}+\frac{%
\left( \Upsilon -\upsilon ^{\prime }\right) }{2}}\mathfrak{N}\left( -\frac{%
k-N_{\Upsilon }-\xi +N_{\upsilon ^{\prime }}-\left( \Upsilon -\upsilon
^{\prime }\right) }{\sqrt{\left( \Upsilon -\upsilon ^{\prime }\right) }}%
\right) \phi \left( \upsilon ^{\prime }\right) d\upsilon ^{\prime } \\ 
+e^{k}\int_{0}^{\Upsilon }\frac{\exp \left( -\frac{\left( k-N_{\Upsilon
}-\xi +N_{\upsilon ^{\prime }}\right) ^{2}}{2\left( \Upsilon -\upsilon
^{\prime }\right) }\right) }{\sqrt{2\pi \left( \Upsilon -\upsilon ^{\prime
}\right) }}\phi \left( \upsilon ^{\prime }\right) d\upsilon ^{\prime }.%
\end{array}
\label{Eq78}
\end{equation}

We found that it more efficient to price these options using the backward
induction. For example, to price the no-touch option backward, we introduce
the new time variable $\varpi =\Upsilon -\upsilon $, and the boundary $%
O_{\varpi }=N_{\Upsilon -\upsilon }$, and write $P\left( 0,T;\xi \right) $
in the form 
\begin{equation}
P\left( 0,T;\xi \right) =e^{-rT}\left( 1-Q\left( \Upsilon ;\xi \right)
\right) .  \label{Eq78a}
\end{equation}%
Here 
\begin{equation}
Q\left( \Upsilon ;\xi \right) =\int_{0}^{\Upsilon }\frac{\left( -\xi
+O_{\varpi ^{\prime }}\right) \exp \left( -\frac{\left( -\xi +O_{\varpi
^{\prime }}\right) ^{2}}{2\left( \Upsilon -\varpi ^{\prime }\right) }\right) 
}{\sqrt{2\pi \left( \Upsilon -\varpi ^{\prime }\right) ^{3}}}\psi ^{\left(
NT\right) }\left( \varpi ^{\prime }\right) d\varpi ^{\prime },  \label{Eq78b}
\end{equation}%
where $\psi ^{\left( NT\right) }\left( \varpi \right) $ solves Eq. (\ref%
{Eq63}) with $f\left( \varpi \right) =1$.

By the same token, we can represent $C\left( 0,S_{0},T,K;\xi \right) $ as
follows%
\begin{equation}
C\left( 0,S_{0},T,K;\xi \right) =e^{-rT}S_{0}\left( D\left( \Upsilon
,k\right) -E\left( \Upsilon ,k;\xi \right) \right) ,  \label{Eq78c}
\end{equation}%
where%
\begin{equation}
\begin{array}{c}
D\left( \Upsilon ,k\right) =e^{k}\left( e^{O_{0}-k+\frac{\Upsilon }{2}}%
\mathfrak{N}\left( \frac{O_{0}-k+\Upsilon }{\sqrt{\Upsilon }}\right) -%
\mathfrak{N}\left( \frac{O_{0}-k}{\sqrt{\Upsilon }}\right) \right) , \\ 
E\left( \Upsilon ,k;\xi \right) =\int_{0}^{\Upsilon }\frac{\left( -\xi
+O_{\varpi ^{\prime }}\right) \exp \left( -\frac{\left( -\xi +O_{\varpi
^{\prime }}\right) ^{2}}{2\left( \Upsilon -\varpi ^{\prime }\right) }\right) 
}{\sqrt{2\pi \left( \Upsilon -\varpi ^{\prime }\right) ^{3}}}\psi ^{\left(
C\right) }\left( \varpi ^{\prime }\right) d\varpi ^{\prime },%
\end{array}
\label{Eq78d}
\end{equation}%
and $\psi ^{\left( C\right) }\left( \varpi ^{\prime }\right) $ solves Eq. (%
\ref{Eq63}) with 
\begin{equation}
\begin{array}{c}
f\left( \varpi \right) =e^{k}\left( e^{\xi -O_{\varpi ^{\prime }}+O_{0}-k+%
\frac{\Upsilon }{2}}\mathfrak{N}\left( \frac{\xi -O_{\varpi ^{\prime
}}+O_{0}-k+\Upsilon }{\sqrt{\Upsilon }}\right) \right. \\ 
\left. -\mathfrak{N}\left( \frac{\xi -O_{\varpi ^{\prime }}+O_{0}-k}{\sqrt{%
\Upsilon }}\right) \right) .%
\end{array}
\label{Eq78e}
\end{equation}

\begin{figure}[tbp]
\begin{center}
\includegraphics[width=0.8\textwidth]
{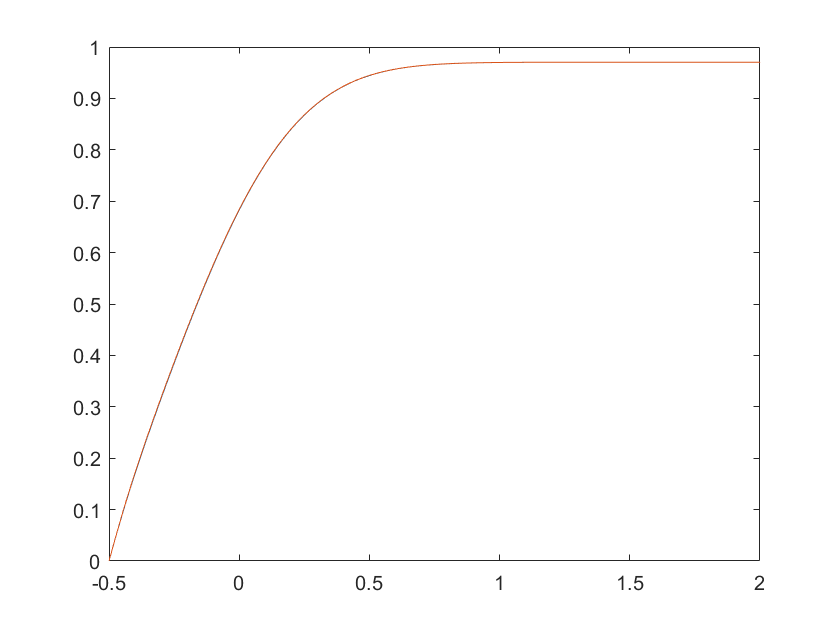}
\end{center}
\caption{$P\left( T,x\right) $ computed via the FDM and the MHP. The
absolute difference between the corresponding functions is less that $%
10^{-3} $. The parameters are the same as in Figure \protect\ref{Fig1}. The
log-barrier is $\protect\xi =-0.5.$}
\label{Fig7}
\end{figure}

\begin{figure}[tbp]
\begin{center}
\includegraphics[width=0.8\textwidth]
{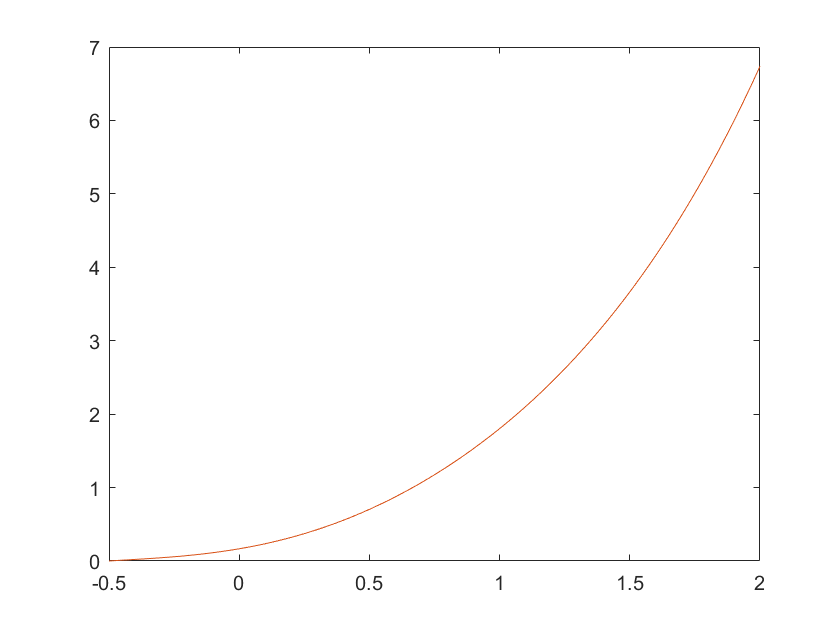}
\end{center}
\par
\vspace{-10pt}
\caption{$C\left( T,x\right) $ computed via the FDM and the MHP. The
absolute difference between the corresponding functions is less that $%
10^{-4} $. The parameters are the same as in Figure \protect\ref{Fig1}. The
log-barrier is $\protect\xi =-0.5.$}
\label{Fig8}
\end{figure}

In Figures \ref{Fig7}, \ref{Fig8} we compare prices of no-touch and call
options obtained via the FDM and the MHP. The figure shows that the
corresponding prices are very close. As before, the MHP is clearly much
faster than the FDM.

\subsection{External loop: averaging over all variance paths\label{Sec43}}

After the corresponding solution is found and expressed in the original
variables, we produce a set of random sequences $\left\{
v_{k}|k=0,1,...,K\right\} $ and repeat steps. Once a sufficiently large
number of paths is generated, we perform averaging and obtain the solution.

\begin{figure}[tbp]
\begin{center}
\subfloat[]{\includegraphics[width=0.75\textwidth]
{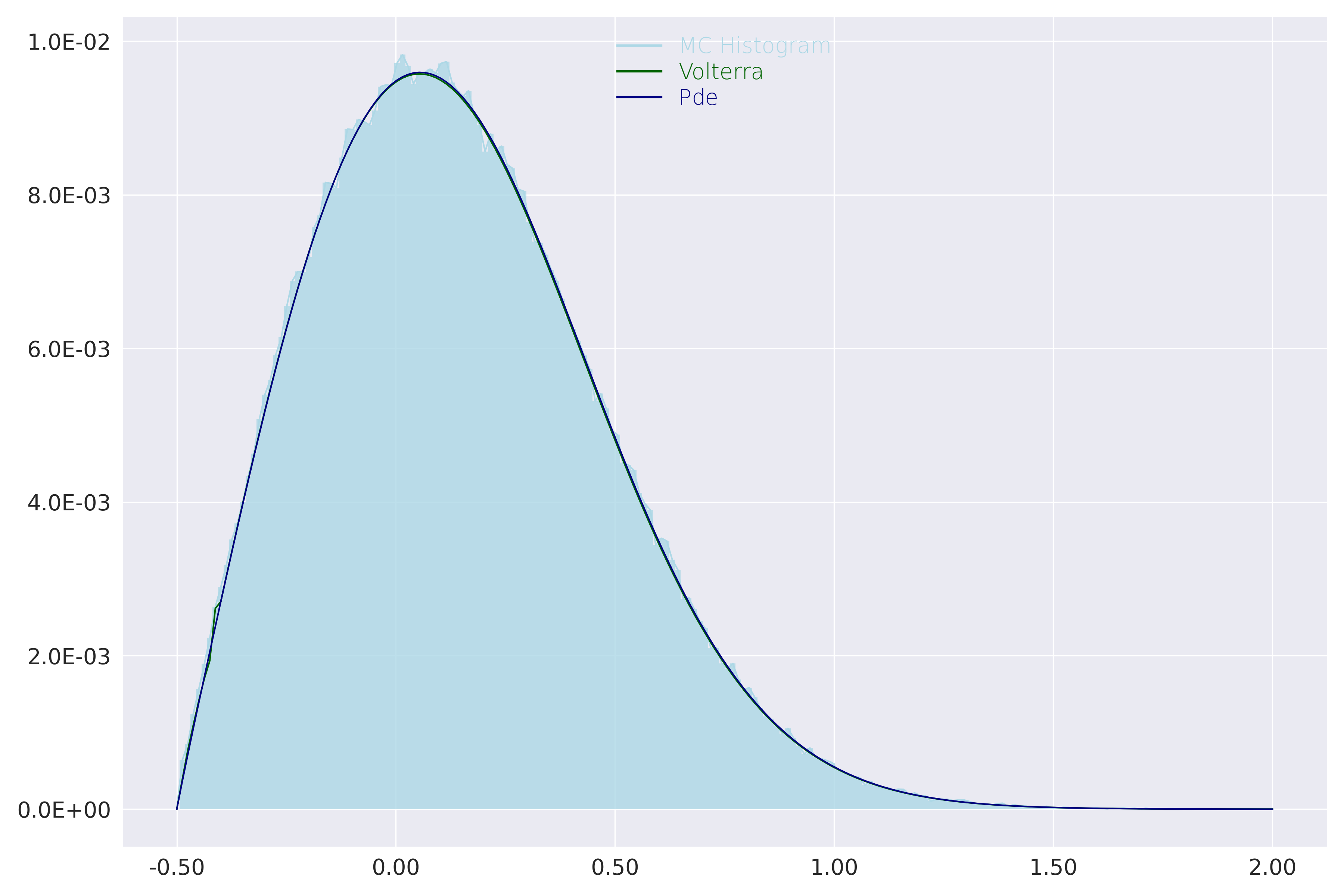}} \\%
[0pt]
\subfloat[]{\includegraphics[width=0.75\textwidth]
{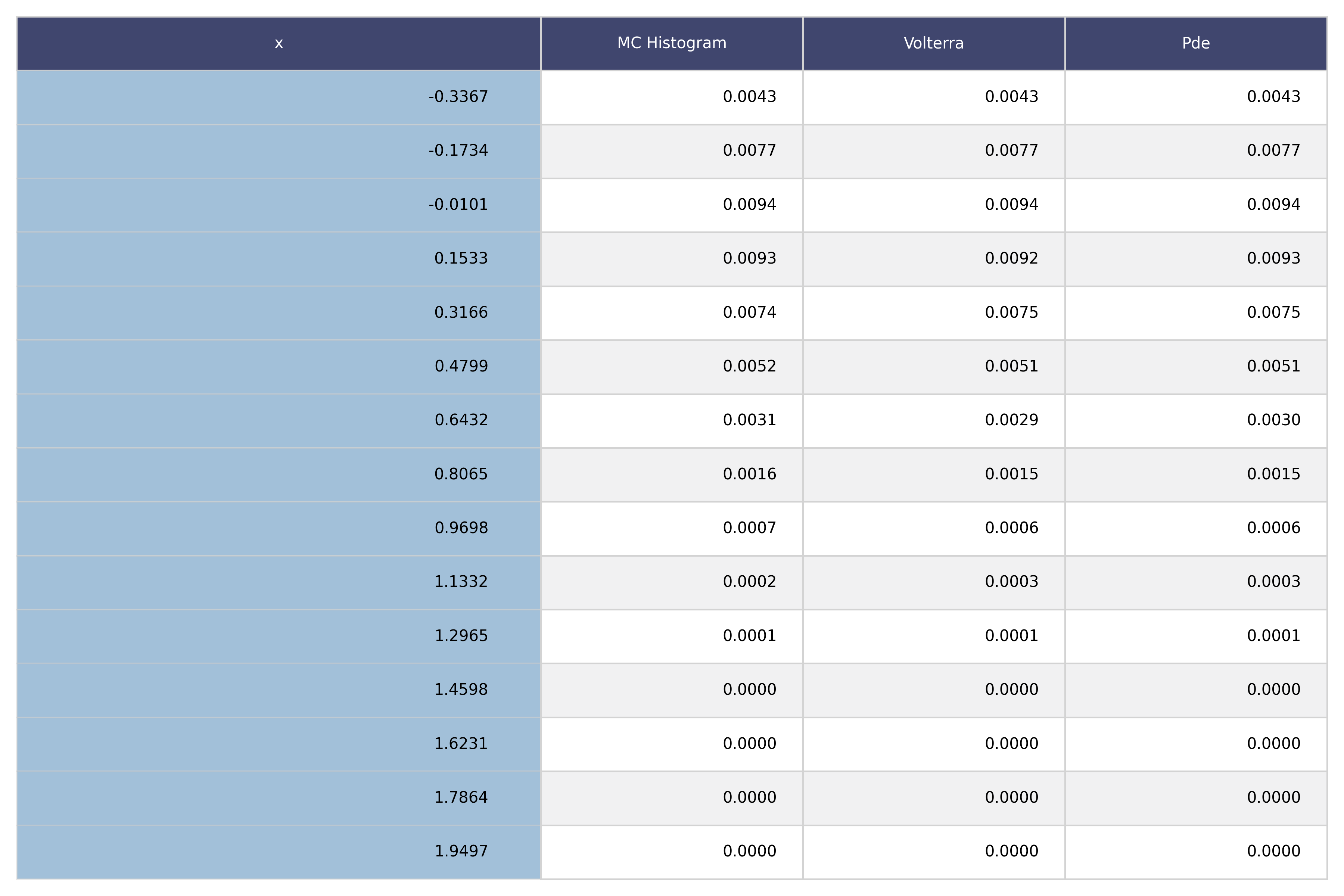}}\\[%
0pt]
\end{center}
\par
\vspace{-10pt}
\caption{Green's function averaged over Monte Carlo paths. The number of MC
path for the MHP and FDM is $10,000$; the number of path for the classical
MCS is $100,000$.}
\label{Fig9}
\end{figure}

\begin{figure}[tbp]
\begin{center}
\subfloat[]{\includegraphics[width=0.75\textwidth]
{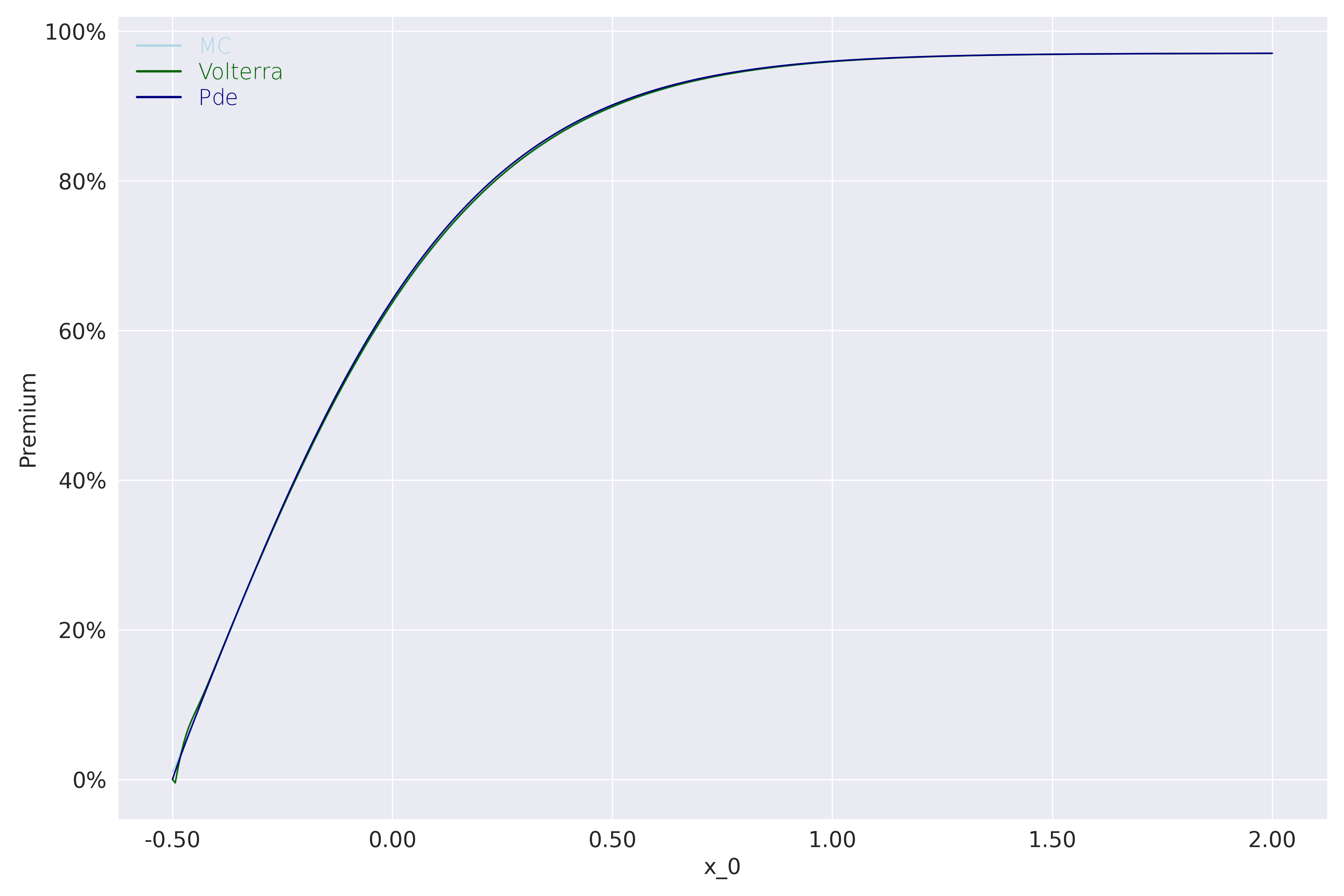}} 
\\[0pt]
\subfloat[]{\includegraphics[width=0.75\textwidth]
{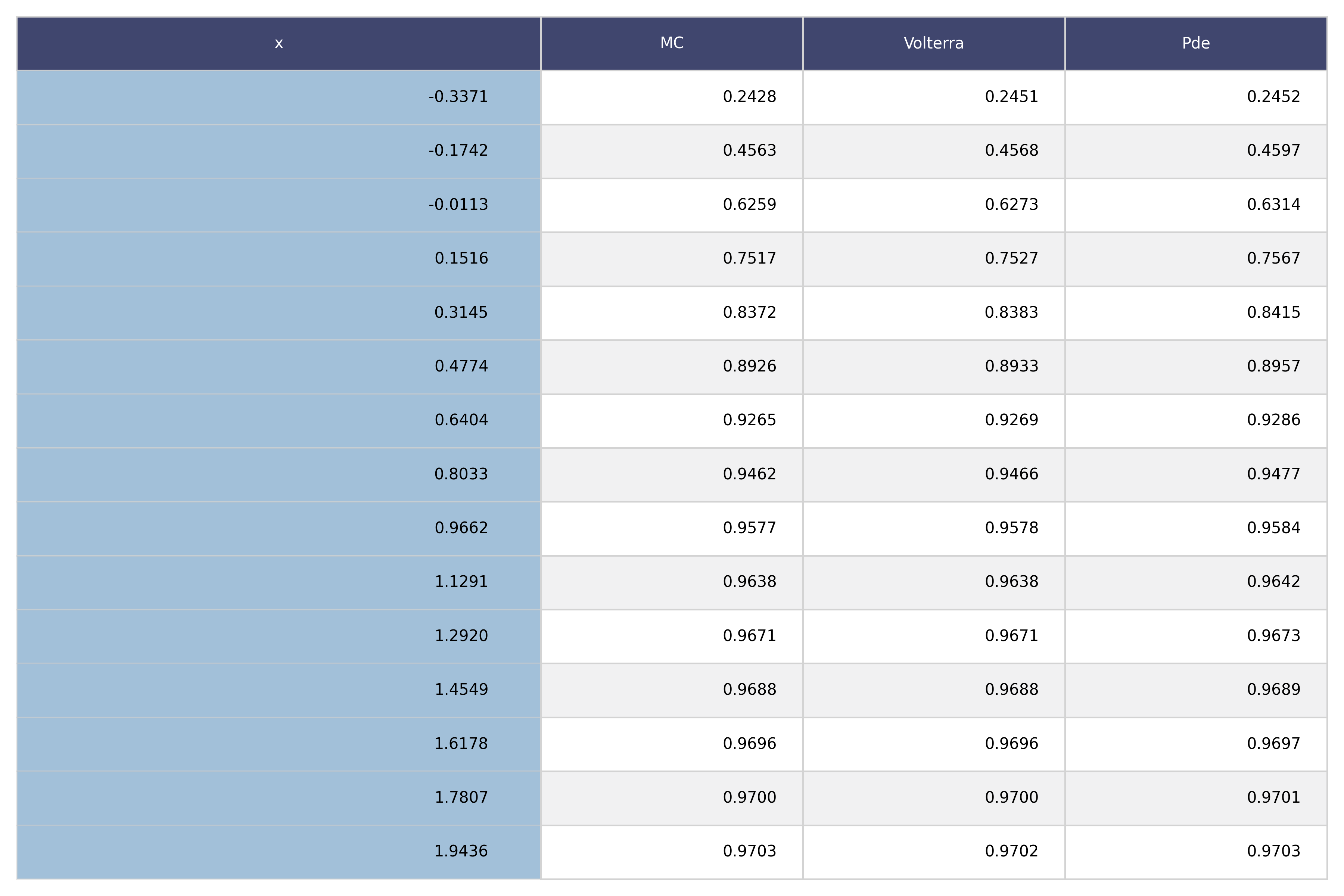}}\\%
[0pt]
\end{center}
\par
\vspace{-10pt}
\caption{Price of the no-touch with the log-barrier = $\protect\xi =-0.5$:
figure (a), table (b). The number of MC paths for the MHP and MFD is $10,000$%
; the number of patha for the MCS is $100,000$.}
\label{Fig10}
\end{figure}

\begin{figure}[tbp]
\begin{center}
\subfloat[]{\includegraphics[width=0.75\textwidth]
{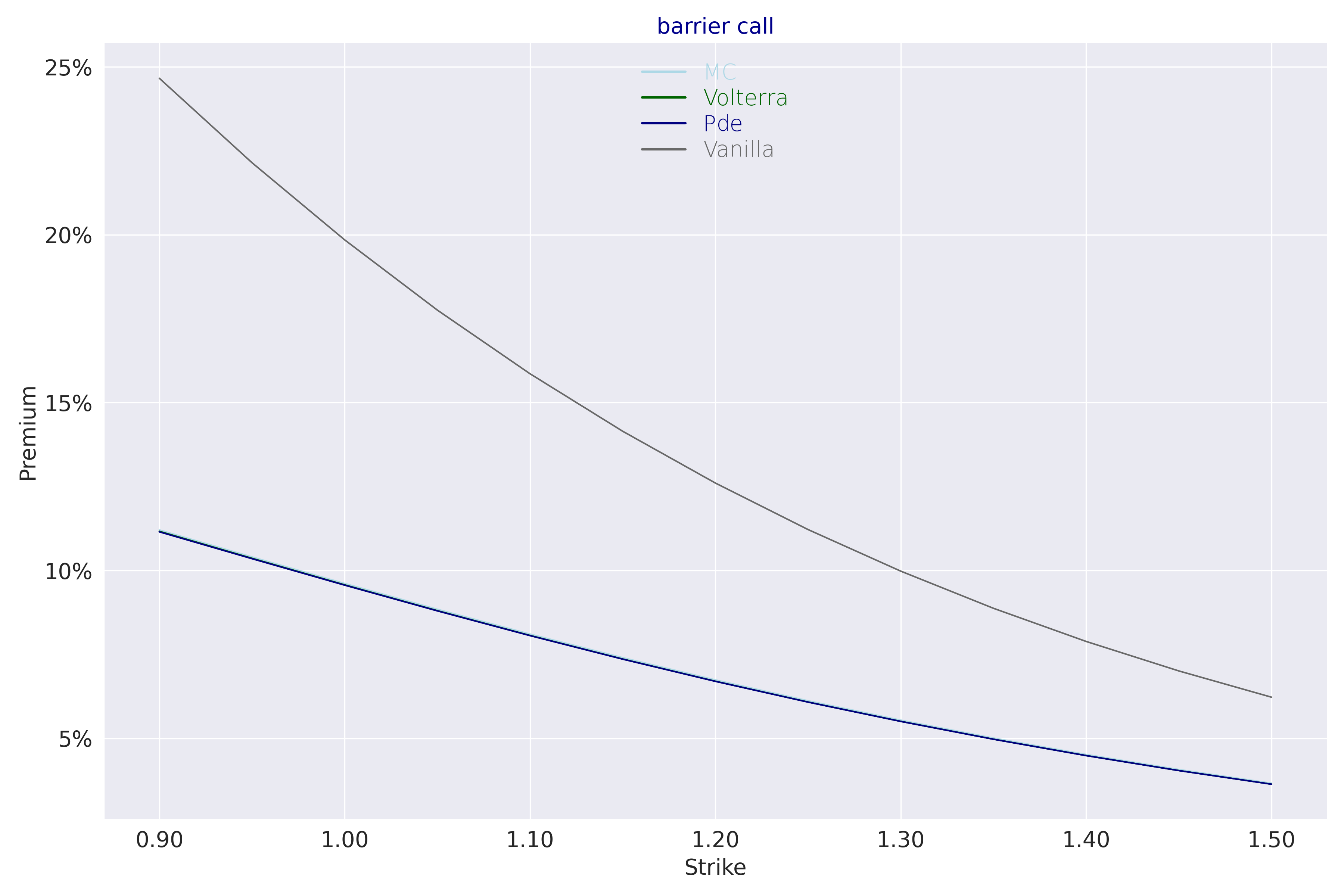}} 
\\[0pt]
\subfloat[]{\includegraphics[width=0.75\textwidth]
{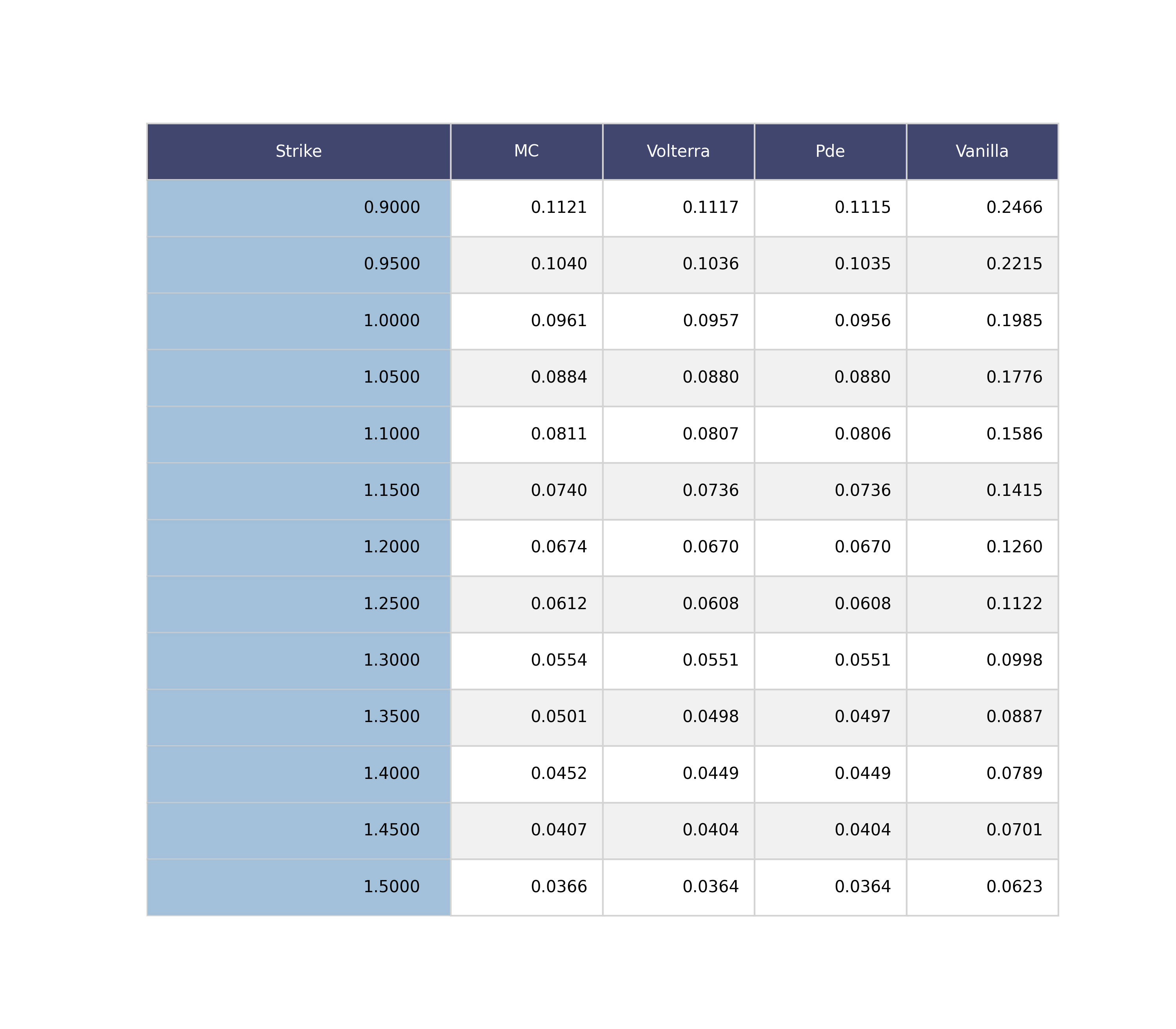}}\\%
[0pt]
\end{center}
\par
\vspace{-10pt}
\caption{(a) The price of the down-and-out call with the barrier at $S=0.9$
as a function of the strike $K$; (b) a summary table. The number of MC paths
for the MHP and MFD is $10,000$; the number of paths for the MCS is $100,000$%
.}
\label{Fig11}
\end{figure}

In Figures \ref{Fig9}, \ref{Fig10}, \ref{Fig11},we show the averaged value
of Green's function, as well as prices of the no-touch and call options.

\section{ Joint probability density for the value of drifted Brownian motion
and its running minimum\label{Sec5}}

This section, which serves as a mathematical aside, shows that the MHP
allows solving a very complex problem of finding the joint probability
distribution for a Brownian motion with time-dependent drift and volatility
and its minimum. Without loss of generality, we can restrict ourselves to
the case of time-dependent drift and unit volatility by scaling time.%
\footnote{%
Despite this being a classical problem, its correct solution is not
presented in the literature, except for the simple case of constant drift.
At the same time, several incorrect solutions have been proposed.}

To put things in perspective, we start with the standard Brownian motion and
consider the following problem:%
\begin{equation}
\begin{array}{c}
G_{t}-\frac{1}{2}G_{xx}=0, \\ 
G\left( 0,x\right) =\delta \left( x\right) ,\ \ \ G(t,a)=0,%
\end{array}
\label{Eq79}
\end{equation}%
where $a$ is the lower bound, $a<0$. It is clear that $G\left( T,b;a\right)
db$ is the probability of the Brownian motion ending in the interval $\left(
b-db/2,b+db/2\right) $ and having its minimum on the interval $\left[ a,0%
\right] $. Thus, the corresponding joint pdf has the form%
\begin{equation}
\pi \left( T,a,b\right) =-\frac{\partial }{\partial a}G\left( T,b;a\right) .
\label{Eq80}
\end{equation}%
The method of images yields%
\begin{equation}
G\left( T,b;a\right) =H\left( T,b\right) -H\left( T,b-2a\right) ,
\label{Eq81}
\end{equation}%
so that%
\begin{equation}
\pi \left( T,a,b\right) =\frac{2}{T}\left( b-2a\right) H\left( T,b-2a\right)
.  \label{Eq82}
\end{equation}

For the drifted Brownian motion the problem can be written as follows:%
\begin{equation}
\begin{array}{c}
G_{t}+\lambda G_{x}-\frac{1}{2}G_{xx}=0, \\ 
G\left( 0,x\right) =\delta \left( x\right) ,\ \ \ G(t,a)=0,%
\end{array}
\label{Eq83}
\end{equation}%
\begin{equation}
G\left( T,b;a\right) =H\left( T,b-\lambda T\right) -e^{2\lambda a}H\left(
T,b-\lambda T-2a\right) ,  \label{Eq84}
\end{equation}%
\begin{equation}
\begin{array}{c}
\pi \left( T,a,b\right) =-\frac{\partial }{\partial a}G\left( T,b;a\right)
\\ 
=\frac{2}{T}\left( b-2a\right) e^{2\lambda a}H\left( T,b-\lambda T-2a\right)
.%
\end{array}
\label{Eq85}
\end{equation}%
Needless to say that for zero drift, $\lambda =0$, we recover Eq. (\ref{Eq82}%
).

Expressions (\ref{Eq82}) and (\ref{Eq85}) are very well-known, even though
their derivation is usually somewhat convoluted. However, to the best of our
knowledge, what is not known, is a similar expression when the drift $%
\lambda $ depends on the (scaled) time $\upsilon $. We shall use the MHP to
derive the corresponding formula. The problem of interest has the form%
\begin{equation}
\begin{array}{c}
G_{\upsilon }+\lambda \left( \upsilon \right) G_{x}-\frac{1}{2}G_{xx}=0, \\ 
G\left( 0,x\right) =\delta \left( x\right) ,\ \ \ G(\upsilon ,a)=0,%
\end{array}
\label{Eq86}
\end{equation}%
\begin{equation}
G\left( \Upsilon ,b;a\right) =H\left( \Upsilon ,b-N_{\Upsilon }\right)
-F\left( \Upsilon ,b;a\right) ,  \label{Eq87}
\end{equation}%
\begin{equation}
F\left( \Upsilon ,b;a\right) =\int_{0}^{\Upsilon }\frac{\left( b+\Theta
\left( \Upsilon ,\upsilon ^{\prime }\right) \left( \Upsilon -\upsilon
^{\prime }\right) \right) \exp \left( -\frac{\left( b+\Theta \left( \Upsilon
,\upsilon ^{\prime }\right) \left( \Upsilon -\upsilon ^{\prime }\right)
\right) ^{2}}{2\left( \Upsilon -\upsilon ^{\prime }\right) }\right) }{\sqrt{%
2\pi \left( \Upsilon -\upsilon ^{\prime }\right) ^{3}}}\phi \left( \upsilon
^{\prime };a\right) d\upsilon ^{\prime }.  \label{Eq88}
\end{equation}%
\begin{equation}
\phi \left( \upsilon ;a\right) +\int_{0}^{\upsilon }\frac{\Theta \left(
\upsilon ,\upsilon ^{\prime }\right) \Xi \left( \upsilon ,\upsilon ^{\prime
}\right) }{\sqrt{2\pi \left( \upsilon -\upsilon ^{\prime }\right) }}\phi
\left( \upsilon ^{\prime };a\right) d\upsilon ^{\prime }=f\left( \upsilon
;a\right) ,  \label{Eq89}
\end{equation}%
\begin{equation}
f\left( \upsilon ;a\right) =H\left( \upsilon ,-N_{\upsilon }-a\right)
=H\left( \upsilon ,N_{\upsilon }+a\right) .  \label{Eq90}
\end{equation}%
Thus,%
\begin{equation}
\begin{array}{c}
\pi \left( \Upsilon ,a,b\right) =-\frac{\partial }{\partial a}G\left(
\Upsilon ,b;a\right) \\ 
=\int_{0}^{\Upsilon }\frac{\left( b+\Theta \left( \Upsilon ,\upsilon
^{\prime }\right) \left( \Upsilon -\upsilon ^{\prime }\right) \right) \exp
\left( -\frac{\left( b+\Theta \left( \Upsilon ,\upsilon ^{\prime }\right)
\left( \Upsilon -\upsilon ^{\prime }\right) \right) ^{2}}{2\left( \Upsilon
-\upsilon ^{\prime }\right) }\right) }{\sqrt{2\pi \left( \Upsilon -\upsilon
^{\prime }\right) ^{3}}}\psi \left( \upsilon ^{\prime };a\right) d\upsilon
^{\prime },%
\end{array}
\label{Eq91}
\end{equation}%
where%
\begin{equation}
\psi \left( \upsilon ;a\right) +\int_{0}^{\upsilon }\frac{\Theta \left(
\upsilon ,\upsilon ^{\prime }\right) \Xi \left( \upsilon ,\upsilon ^{\prime
}\right) }{\sqrt{2\pi \left( \upsilon -\upsilon ^{\prime }\right) }}\psi
\left( \upsilon ^{\prime };a\right) d\upsilon ^{\prime }=\frac{\left(
N_{\upsilon }+a\right) }{\upsilon }f\left( \upsilon ,a\right) .  \label{Eq92}
\end{equation}

\section{Conclusions\label{Sec6}}

This paper introduced a new hybrid MHP/MCS technique for pricing barrier
options on assets with stochastic volatility. The idea is to decompose the
solution process into the inner step, which solves a barrier problem for the
conditionally independent process, and the outer step, which averages the
corresponding solutions over the one-dimensional stochastic volatility
dynamics.

Our methodology is general and can manage all known stochastic volatility
models equally efficiently. Besides, relatively simple extensions (which
will be described elsewhere) can also handle rough volatility models. With
minimal changes, one can use the method to price popular double-no-touch
options and other similar instruments.

While several authors used hybrid techniques before, see, e.g., \cite%
{Loeper2009}, their methods use the FDM and are still relatively slow,
although undeniably faster than the standard two-dimensional MCS. Our method
reduces the inner barrier problem to solving a linear Volterra equation of
the second kind. It is very efficient and is an order of magnitude faster
than other hybrid methods with the same (or better) accuracy. Our results
are a natural generalization of Willard's formula, see \cite{Willard1997},
for barrier options.

As a byproduct of our analysis, we derived a new expression for the joint
pdf for the value of a drifted Brownian motion and its running minimum or
maximum in the case of time-dependent drift.

\begin{acknowledgement}
We are grateful to Andrey Itkin and Dmitry Muravey for several useful
discussions of the topics covered in this paper. Numerous conversations with
Marcos Lopez de Prado and Alexey Kondratiev were very helpful.
\end{acknowledgement}

\end{document}